\documentclass[two column,showkeys]{revtex4}
\pdfoutput=1
\usepackage[T1]{fontenc}
\usepackage{graphicx}
\usepackage{amsmath}
\usepackage{amssymb}
\usepackage{amsfonts}
\usepackage{color}
\usepackage{appendix}

\def\a{\alpha}

\def\g{\gamma}
\def\e{\varepsilon}
\def\d{\delta}

\def\k{\kappa}

\def\m{\mu}

\def\t{\tau}

\def\g{\gamma}
\def\r{\rho}
\def\s{\sigma}

\def\S{\Sigma}
\def\G{\Gamma}
\def\D{\Delta}
\def\o{\omega}

\def\be{\begin{equation}}
\def\ee{\end{equation}}
\def\bea{\begin{eqnarray}}
\def\eea{\end{eqnarray}}
\def\llang{\langle\langle}
\def\rrang{\rangle\rangle}
\def\nn{\nonumber}
\def\lb{\label}

\begin{document}
	
	\title{Topological conditions for impurity effects in carbon nanosystems}
	
	\author{Yuriy G. Pogorelov}%
	\email{ypogorel@fc.up.pt}
	\affiliation{IFIMUP-IN,~Departamento~de~F\'{i}sica,~Universidade~do~Porto,~Porto,~Portugal,}
	
	\author{Vadim M. Loktev}%
	\email{vloktev@bitp.kiev.ua}
	\affiliation{N.~N.~Bogolyubov~Institute~of~Theoretical~Physics,~NAS~of~Ukraine,~Kyiv,~Ukraine, \\
		\&\\
		Igor~Sikorsky~Kyiv~Polytechnic~Institute,~Kyiv,~Ukraine,}
	
	\begin{abstract}
		We consider electronic spectra of carbon nanotubes and their perturbation by impurity atoms absorbed at different positions on nanotube 
		surfaces, within the framework of Anderson hybrid model. A special attention is given to the cases when Dirac-like 1D modes appear in the 
		nanotube spectrum and their hybridization with localized impurity states produces, at growing impurity concentration $c$, onset of a mobility 
		gap near the impurity level and even opening, at yet higher $c$, of some narrow delocalized range within this mobility gap. Such behaviors are 
		compared with the similar effects in the previously studied 2D graphene and armchair type graphene nanoribbons. Some possible practical 
		applications are discussed.
	\end{abstract}

	\date{\today}
	\maketitle

	\section{Introduction}\lb{intr}
	
	From the discovery, 20 years ago, of single-layer graphene \cite{Geim2004}, an enormous interest has been attracted not only to its two-dimensionality 
	(2D) \cite{Geim2005} but mainly to its massless, that is Dirac-like, spectrum of electronic excitations \cite{Geim2009}. Studies of its various 
	physical properties have a very broad nomenclature \cite{Tomanek_Book, Inagaki2014} (see also \cite{Katsnelson}) but we focus here on certain aspects 
	of non-ideal graphene structures, yet restricted to a single dimension (1D), namely, of graphene nanoribbons (NR's) \cite{Wakabayashi2010} and carbon
	nanotubes (NT's) \cite{Iijima2002, Charlier2007} in presence of impurities \cite{Nevidomskyy, Jalili, Pumera, Skrypnyk, Vejpravova}. Mostly, we consider here 
	the electron quasiparticle spectra in principal topological types of graphene 1D nanosystems and their restructuring under effects by impurity atoms 
	absorbed at different positions over carbon atoms \cite{Wehling2009, Araujo}. In this course, the main attention is given to the cases when Dirac-like 
	1D modes are present in the NT spectrum \cite{Wakabayashi1996} and we compare the impurity disorder effects on such modes with those previously 
	studied in 2D graphene \cite{YDV2021} and in armchair type nanoribbons (ANR's) \cite{PL2022}. Also, the specifics of impurity effects in more general 
	twisted carbon NT's are shortly discussed. The main purpose of this analysis is in finding possibile practical applications for such 1D-like semi-metallic 
	systems under thier properly adjusted doping as more compact and sensible analogues for common doped semiconductors. 
	
	The presentation is organized as follows. We begin from description of quasiparticle spectra for two basic NT topologies: zigzag (ZNT's, 
	Sec. \ref{Zigzag}) and armchair (ANT's, Sec. \ref{Arm}), in the forms adjusted to describe the impurity induced restructuring of their 
	spectra. This description, within the simplest T-matrix approximation for the quasi-particle self-energy, begins from the technically simpler ANT case 
	(Sec. \ref{Imp}) and then extends to a more involved ZNT case (Sec. \ref{ZNT}). The next comparison with the previous results for 2D graphene and ANR's 
	(Sec. \ref{Comp}) reveals both qualitative similarities and some quantitative differences in their behaviors. At least, the general topology of twisted 
	nanotubes (TNT's) is discussed in Sec. \ref{TNT}, suggesting a qualitative difference between the twisted and non-twisted NT's in their sensitivity to 
	impurity disorder. The obtained results are then verified with some T-matrix improvements (Sec. \ref{Beyond}): the self-consistent T-matrix method and the 
	group expansion (GE) method, both of them confirming validity of the simple T-matrix picture. The final discussion of these theoretical results and of some 
	perspectives for their practical applications is given in Sec. \ref{Disc}.
	
	\section{Zigzag nanotubes}\lb{Zigzag} 
	Carbon NT's can be obtained from carbon NR's by closure of their edges (for instance, of basic zigzag or armchair types), and these nanotubes are usually 
	classified by the normals to their axes (that is to the related NR edges). Thus, folding of an armchair nanoribbon (ANR) produces a ZNT and, vice versa, 
	that of a zigzag nanoribbon (ZNR) does an ANT.
	
\begin{figure}[h!]
		\centering
		\includegraphics[width=9.5cm]{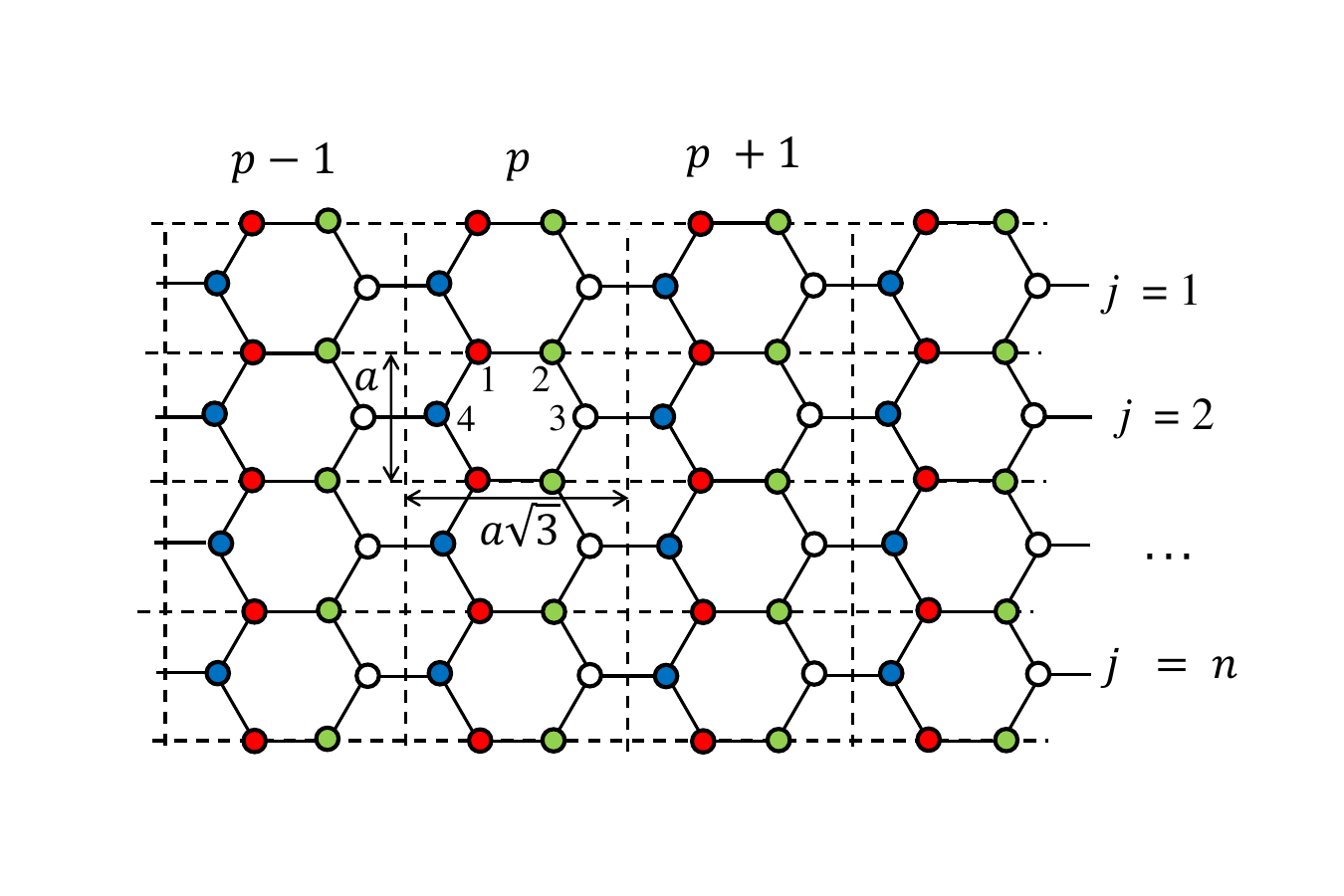}
		\caption{An armchair nanoribbon composed of $n$ atomic chains ($j$-labeled), each chain consisting of segments ($p$-labeled) with 4 atomic sites: 
			$s = 1$ (red), $2$ (green), $3$ (white), $4$ (blue), in each segment.}
		\label{fig1}
	\end{figure}

	Beginning from the ANR case, it can be seen as a composite of $n$ chains (labeled by $j$ indices) of transversal period $a$ (the graphene lattice constant), 
	each chain containing $N \gg 1$ segments (labeled by $p$ indices) of longitudinal period $a\sqrt 3$ and each segment including 4 atomic sites (labeled by 
	$s$ indices, see Fig. \ref{fig1}).
	Next, the closure between the 1st and $n$th chains of an ANR, transforms it into a ZNT (see Fig. \ref{fig2}). 
	\begin{figure}[h!]
		\centering
		\includegraphics[width=8cm]{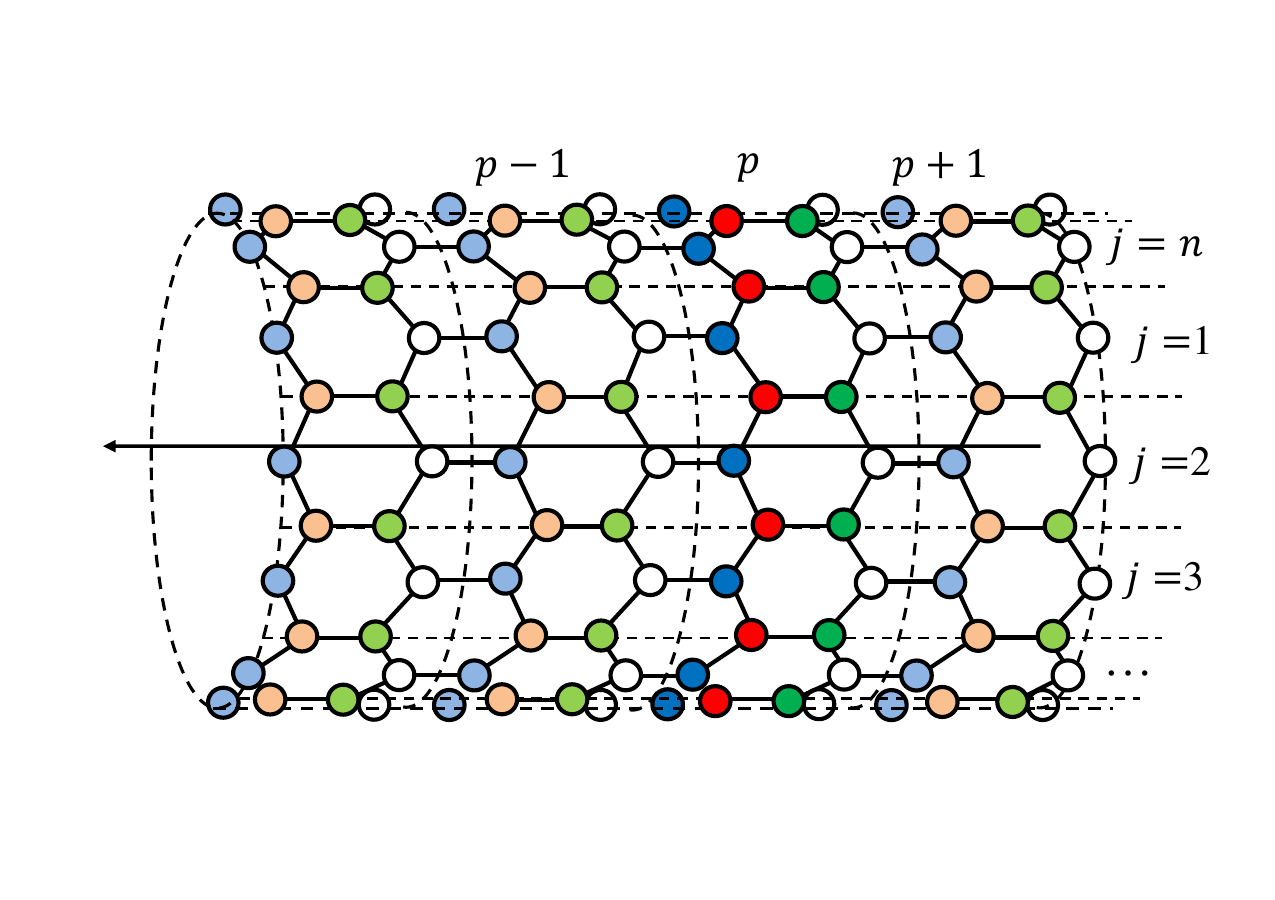}
		\caption{A zigzag nanotube formed by closing links between the 1st and $n$th chains of the armchair nanoribbon in Fig. \ref{fig1}.}
		\lb{fig2}
	\end{figure}

	For the following consideration of electronic dynamics, it is suitable to combine the local Fermi operators $a_{p,j,s}$ at 4 $s$-sites from $j$th chain 
	in $p$th segment into the 4-spinor:
	\be
	a_{p,j} = \left(\begin{array}{c}
		a_{p,j;1}  \\
		a_{p,j;2}  \\
		a_{p,j;3}  \\
		a_{p,j;4}  \\
	\end{array}\right).
	\lb{asp}
	\ee
	Then the longitudinal translation invariance (with the $a\sqrt 3$ period, see Fig. \ref{fig1}) and the discrete transversal rotation invariance of the 
	obtained ZNT suggest the Fourier expansion of the local spinor components in quasi-continuous longitudinal momentum $-\pi/\sqrt 3 < k < \pi/\sqrt 3$ and 
	in discrete transversal wave number $q = 0,\dots, n - 1$ (both in $a^{-1}$ units):
	\be
	a_{p,j,s} = \frac 1{2\sqrt{nN}}\sum_{k,q}\exp\left[i\left(\sqrt 3 kp_s + \frac {2\pi q}n j_s\right)\right]
	\a_{k,q,s}.
	\lb{apj}
	\ee
	Its components $\a_{k,q,s}$ form the wave spinor $\a_{k,q}$. Here the longitudinal numbers for different $s$-sites are: $p_{1,2} = p \pm 1/6$, $p_{3,4} 
	= p \pm 1/3$ and their transversal numbers are: $j_{1,2} = j$, $j_{3,4} = j + 1/2$. Then the ZNT Hamiltonian with only account of hopping between nearest 
	neighbor atoms (its parameter, $t \approx 2.8$ eV \cite{Geim2009}, taken as the energy scale in what follows) is presented in terms of wave spinors as 
	\footnote{The common nearest neighbor hopping approximation stays practically insensible to the NT curvature at $n \gg 1$ since the distance 
		to next-nearest neighbors there stays, within to $\sim 1/n^2$, the same as in 2D graphene.}:
	\be
	H_{ZNT} = \sum_{k,q} \a_{k,q}^\dagger\hat H_{k,q}\a_{k,q}.
	\lb{HZ}
	\ee
	Here the $4\times 4$ matrix:
	\be
	\hat H_{k,q} = \left(\begin{array}{cccc}
		0 & h_k & 0 & h_{k,q}^\ast\\
		h_k^\ast & 0 & h_{k,q} & 0 \\
		0 & h_{k,q}^\ast & 0 & h_k \\
		h_{k,q} & 0 & h_k^\ast & 0
	\end{array}\right)
	\lb{Hkq}
	\ee
	has its elements $h_k = {\rm e}^{i k/\sqrt 3}$ and $h_{k,q} = 2{\rm e}^{i k/2\sqrt 3}\cos\tfrac{\pi q}n$. The ZNT spectrum results from four eigen-values 
	of this matrix at given $k$ and $q$ as: 
	\bea
	&& \e_{k,q;1} = -\e_{k,q;2} = -\e_{k,q},\nn\\
	&& \qquad \e_{k,q;3} = -\e_{k,q;4} = -\e_{k,n -q},
	\lb{ekqf}
	\eea
	with the basic dispersion law:
	\be
	\e_{k,q} = \sqrt{1 + 4\cos \tfrac{\sqrt 3 k}2\cos\tfrac{\pi q}n + 4\cos^2\tfrac {\pi q}n}.
	\lb{ekq}
	\ee
	
	This can be seen either as the standard graphene dispersion law \cite{Wallace} but with discrete transversal momentum numbers $q$ or, otherwise, as a set 
	of $4n$ 1D $k$-bands $\e_{k,q;f}$ (for $n$ possible values of $q$ and 4 values of $f$). Notably, a double degeneracy of these bands follows from Eqs. 
	\ref{ekqf}, \ref{ekq} as:
	\be
	\e_{k,q;1} \equiv \e_{k,n - q;3}, \quad \e_{k,q;2} \equiv \e_{k,n - q;4}.
	\lb{deg}
	\ee
	
	Thus, if $n$ is even, the ZNT spectrum has 4 non-degenerated (for $q = 0$ and $q = n/2$) modes and $2n - 2$ doubly-degenerated ones. Otherwise, if $n$ is odd, 
	there are two non-degenerated modes (only for $q = 0$) and $2n - 1$ doubly-degenerated ones. The eigen-operators $\psi_{k,q;f}$ of these modes enter the 
	diagonal ANT Hamiltonian:
	\be
	H = \sum_{k,q,f}\e_{k,q;f}\psi_{k,q;f}^\dagger\psi_{k,q;f}.
	\lb{Hd}
	\ee
	These operators at given $k$ and $q$ can be also combined into the 4-spinor $\psi_{k,q}$, related to the $\a$-spinor as:
	\be
	\psi_{k,q} = \hat U_{k,q}\a_{k,q},
	\lb{psi}
	\ee
	through the unitary matrix: 
	\[\hat U_{k,q} = \frac 12\left(\begin{array}{cccc}
	-z_{k,q} & -1 & z_{k,q} & 1\\
	z_{k,q} & -1 & -z_{k,q} & 1\\
	-z_{k,n - q} & 1 & -z_{k,n - q} & 1\\
	z_{k,n - q} & 1 & z_{k,n - q} & 1
	\end{array}\right),\]
	with the complex phase factor
	\[z_{k,q} = \exp\left[i\left(\frac k{\sqrt 3} + \arctan \frac{\sin \tfrac k{2\sqrt 3}}{2\cos\tfrac{\pi q}n - \cos\tfrac k{2\sqrt 3}}\right)\right].\]
	Contrariwise, the $\a$-spinor follows from the $\psi$-spinor by the inversion of Eq. \ref{psi}: 
	\be
	\a_{k,q} = \hat U_{k,q}^\dagger \psi_{k,q}.
	\lb{apsi}
	\ee
	
	Another important features of the ZNT spectrum by Eq. \ref{ekqf} are: 
	
	i) presence of 2 flat (dispersionless) modes for even $n$ (then at $q = n/2$) and
	
	ii) presence of 4 gapless Dirac-like modes (DLM's) for $n$ being a multiple of 3 (then at $q = n/3$ and $q = 2n/3$). 
	
	The latter are just the 1D analogs to the 2D graphene Dirac modes with their nodal points $K$ (here at $k = 2\pi/\sqrt 3,\, q = n/3$) and $K'$ (here 
	at $k = 0,\,q = 2n/3$), also they are fully analogous to DLM's in ANR \cite{PL2022}.

	Due to the DLM's special sensitivity to local impurity perturbations, our following treatment is mainly focused on these modes. In this course, the most 
	relevant energy range is the Dirac window (DW), exclusively occupied with DLM's and delimited by the inner edges of their nearest neighbor modes (Fig. \ref{figa}). 
	From Eq. \ref{ekq}, this window results of width:
	\be
	\D_{\rm DW} = 2\left|1 - \cos\frac\pi n - \sqrt 3 \sin\frac\pi n\right|,
	\lb{DW}
	\ee
	and with growing $n \gg 1$ it is narrowing as $\sim 2\sqrt 3\pi/n$.
		\begin{figure}[h]
		\centering
		\includegraphics[width=7cm]{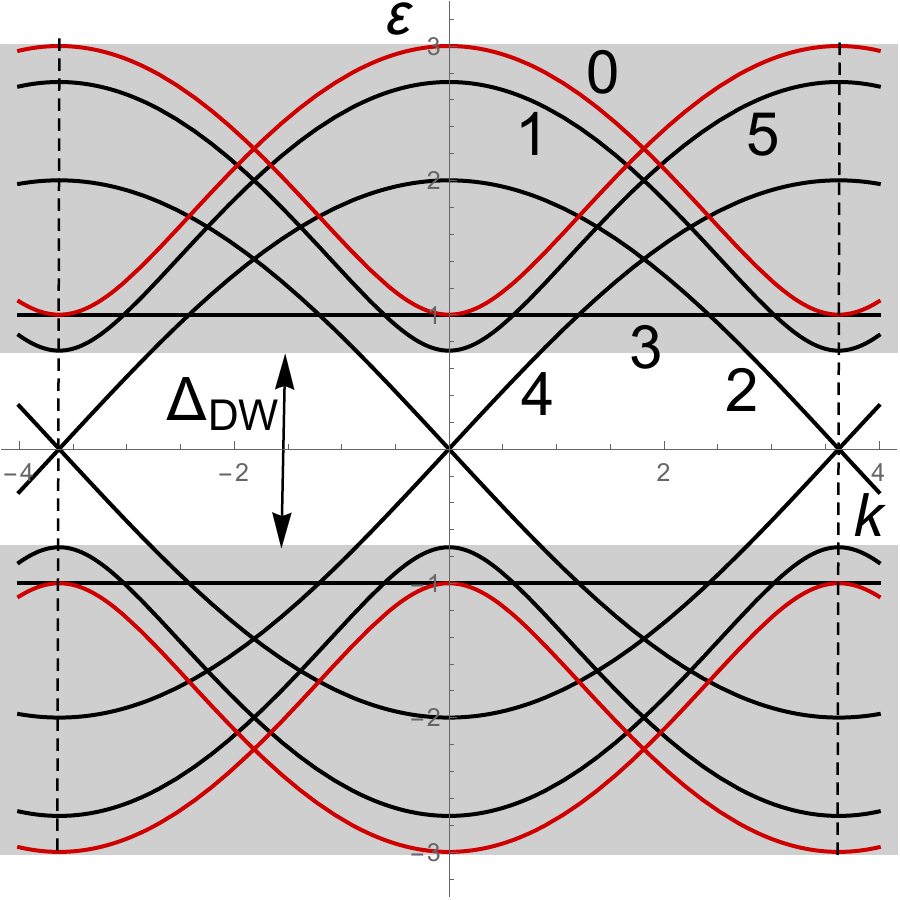}
		\caption{Dispersion laws for ZNT of $n = 6$ chains, with doubly-degenerated (black) and non-degenerated (red) modes, the Dirac window of width 
			$\D_{\rm DW} = 2(\sqrt 3 -1)$ is in between the shaded ranges of the resting modes.}
		\lb{figa}
	\end{figure}
	
	Now, considering the low-energy spectrum range, the expansion of local operators by Eqs. \ref{apj}, \ref{apsi} can be restricted to 8 $K,K'$ DLM's which 
	share 4 eigen-energies: 
	\bea
	&& \e_{k,K,1} = - \e_{k,K,2} = - \e_{k,K',1}\nn\\
	&&\qquad\qquad\qquad = \e_{k,K',2} = 2\sin \frac {\sqrt 3 k}4,\nn\\
	&&\e_{k,K,3} = - \e_{k,K,4} = - \e_{k,K',3}\nn\\
	&&\qquad\qquad\qquad = \e_{k,K',4} = 2\cos \frac {\sqrt 3 k}4.
	\lb{KKa}
	\eea
	
	Thus, the restricted expansion of a local spinor in eigen-spinors is presented in the form:
	\bea
	&& a_{p,j} = \frac 1{2\sqrt{nN}} \sum_k\, {\rm e}^{i\sqrt 3 kp}\left({\rm e}^{i\tfrac{2\pi}3 j}U_{k,K}^\dagger\psi_{k,K}\right.\nn\\
	&&\qquad\qquad\qquad\qquad + \left.{\rm e}^{i\tfrac{4\pi}3 j}U_{k,K'}^\dagger\psi_{k,K'}\right),
	\lb{aKK}
	\eea
	including the unitary matrices:
	\bea
	\hat U_{k,K} & = & \frac 12\left(\begin{array}{cccc}
		-i z_k & -1 & i z_k & 1 \\
		i z_k & -1 & -i z_k & 1 \\
		z_k & 1 &  z_k & 1 \\
		-z_k &  1 & -z_k & 1
	\end{array}\right),\nn\\
	\hat U_{k,K'} & = & \frac 12\left(\begin{array}{cccc}
		- z_k & -1 & z_k & 1 \\
		z_k & -1 & -z_k & 1 \\
		- i z_k & 1 & - i z_k & 1 \\
		iz_k & 1 & iz_k & 1 
	\end{array}\right)\nn
	\eea
	with $z_k = {\rm e}^{ik/2\sqrt 3}$. This expansion is suitable for the following construction of impurity perturbation Hamiltonian.  
	
	\section{Armchair nanotubes}\lb{Arm}
	For the case of ANT we also consider its structure obtained from an $n$-chain ZNR (Fig. \ref{fig5}) by closure between its 1st and $n$th chains 
	(Fig. \ref{fig6}). 
	\begin{figure}[h]
		\centering
		\includegraphics[width=9.5cm]{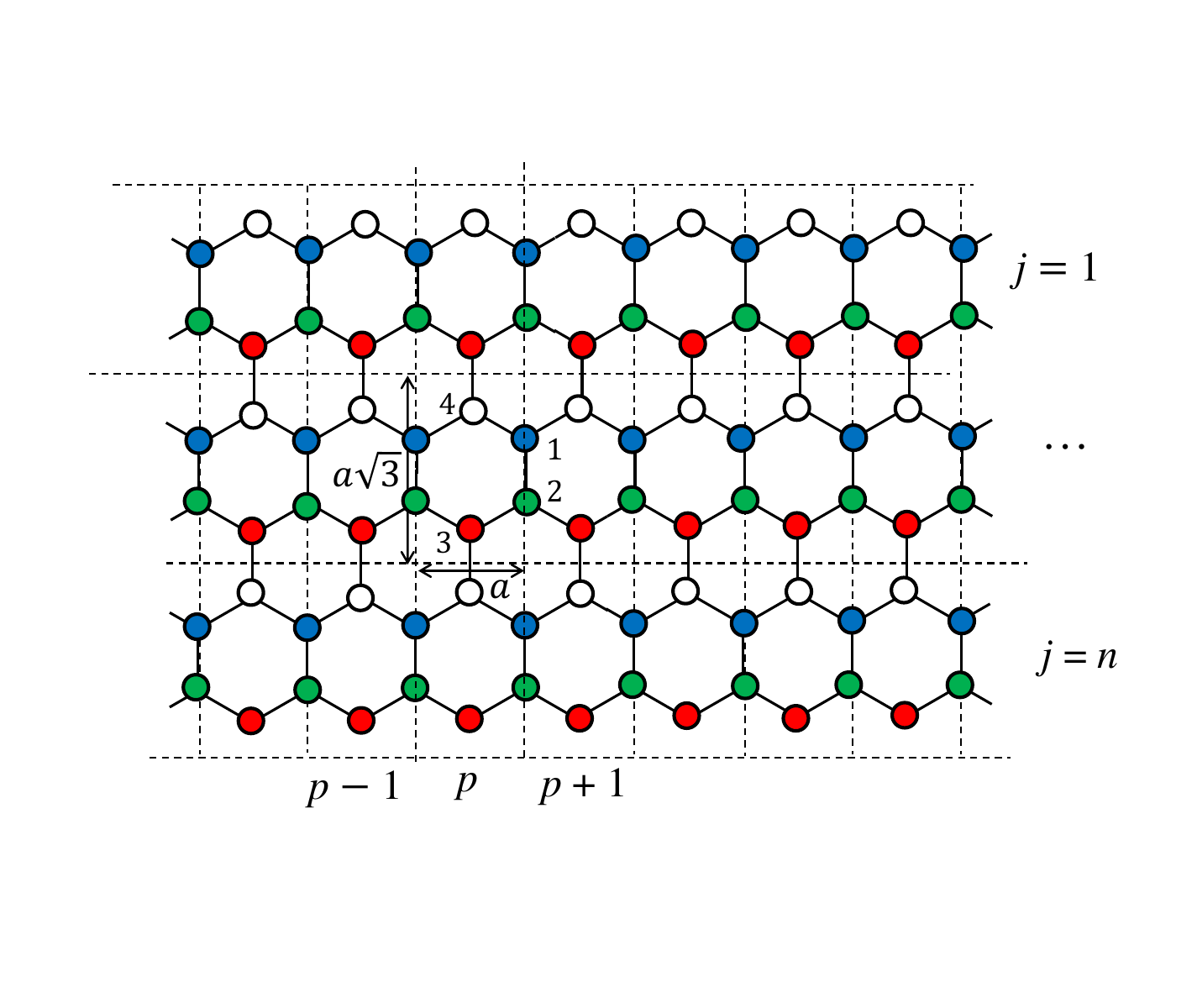}
		\caption{A zigzag nanoribbon of $j = 1, \dots,n$ chains, with $s=1$ (blue), 2 (green), 3 (red), and 4 (white) sites in each $p$th segment.}
		\lb{fig5}
	\end{figure}
	\begin{figure}[h]
		\centering  
		\includegraphics[width=9cm]{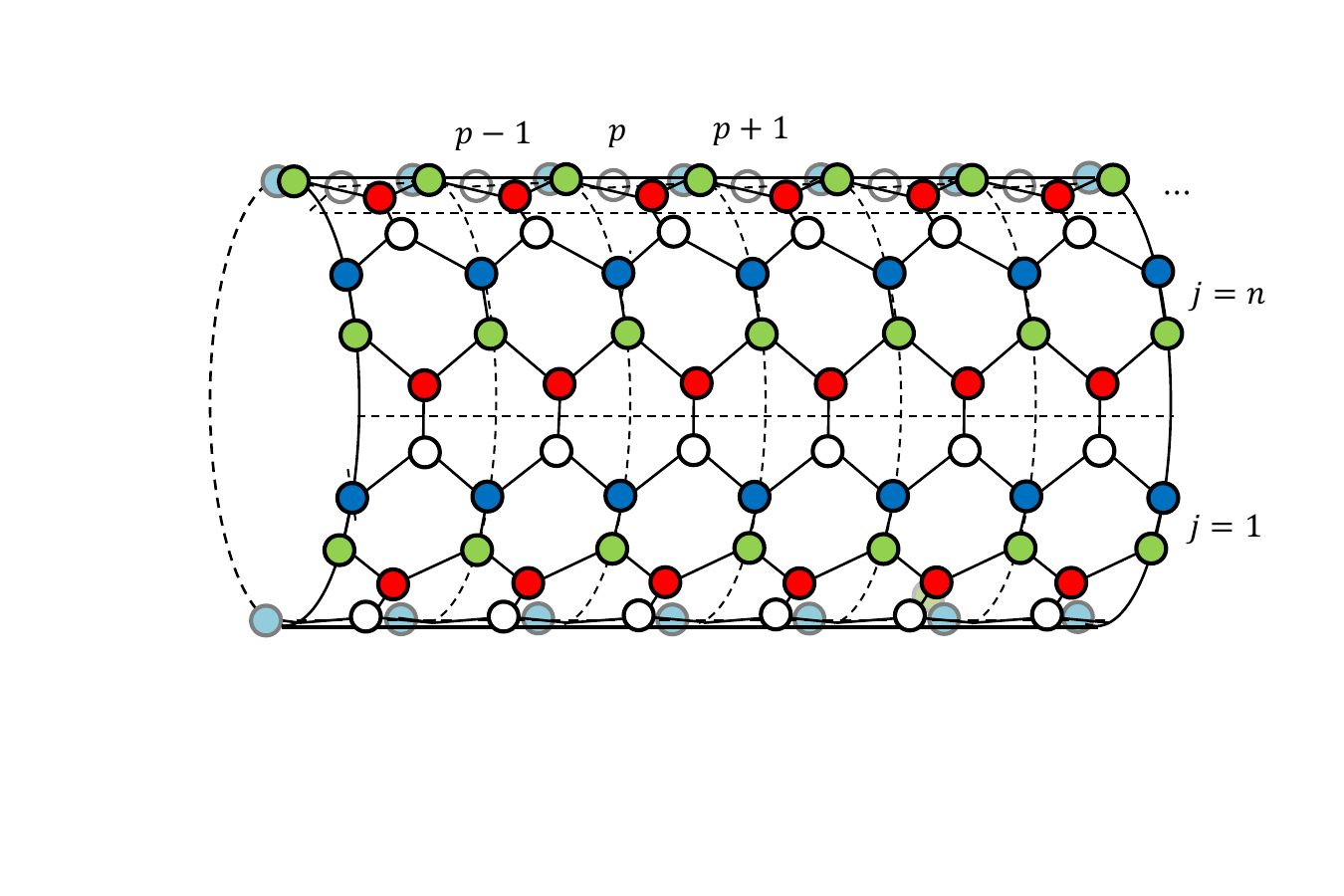}
		\caption{An $n$-chain armchair nanotube formed by closure between the 1st and $n$th chains of zigzag nanoribbon from Fig. \ref{fig5}.}
		\label{fig6}
	\end{figure}
	Comparison of Fig. \ref{fig5} with Fig.  \ref{fig1} readily shows that the ANT elementary cell results just from 90 degrees rotation of the ZNT one, so 
	the analysis of ANT spectra simply follows that for ZNT but with the $\sqrt 3 k \longleftrightarrow 2\pi q/n$ interchange. Thus the ANT Hamiltonian in 
	terms of 4-spinor wave operators results, in analogy with the ZNT form by Eq. \ref{apj}, as:
	\be
	H_{ANT} = \sum_{k,q}\a_{k,q}^\dagger\hat H_{q,k}\a_{k,q},
	\lb{HA}
	\ee
	where the $4\times 4$ matrix:
	\be
	\hat H_{q,k} = \left(\begin{array}{cccc}
		0 & h_q & 0 & h_{q,k}^\ast\\
		h_q^\ast & 0 & h_{q,k} & 0 \\
		0 & h_{q,k}^\ast & 0 & h_q \\
		h_{q,k} & 0 & h_q^\ast & 0
	\end{array}\right)
	\lb{Hzkq}
	\ee
	has its elements $h_q = {\rm e}^{i 2\pi q/3n}$ and $h_{q,k} = 2{\rm e}^{i\pi q/3n}\cos k/2$. The ANT spectrum at given $k$ and $q$ results from the 
	indicated interchange in Eqs. \ref{ekqf}, \ref{ekq}, as: 
	\bea
	&& \e_{q,k;1} = -\e_{q,k;2} = -\e_{q,k},\nn\\
	&& \qquad \e_{q,k;3} = -\e_{q,k;4} = -\e_{n -q,k},
	\lb{zkqf}
	\eea
	where:
	\be
	\e_{q,k} = \sqrt{1 + 4\cos\tfrac{\pi q}n \cos \tfrac k2 + 4\cos^2\tfrac k2}.
	\lb{zkq}
	\ee
	This spectrum includes the same numbers of non-degenerated and doubly degenerated eigen-modes as in the above considered ZNT case. But it differs from 
	that case by: 
	
	i) absence of flat modes and 
	
	ii) presence of two Dirac nodal points: $k = 2\pi/3$ ($K$) and $k = -2\pi/3$ ($K'$) at the same $q = 0$ and for {\it any} ANT width $n$ value. Notably, 
	the related two DLM's are non-degenerated (compare with Eq. \ref{KKa}): 
	\be
	\e_{0,k;3} = -\e_{0,k,4} \equiv \e_k = 1 - 2\cos \tfrac k2. 
	\lb{KKz}
	\ee
	
	For the ANT case, the DW width results $\D_{\rm DW} = 2\sin \pi/n$, that is narrowing with $n \gg 1$ as $\D_{\rm DLM} \sim 2\pi/n$ (to be compared with 
	the ZNT case by Eq. \ref{DW}).
	\begin{figure}[h]
		\centering
		\includegraphics[width=7cm]{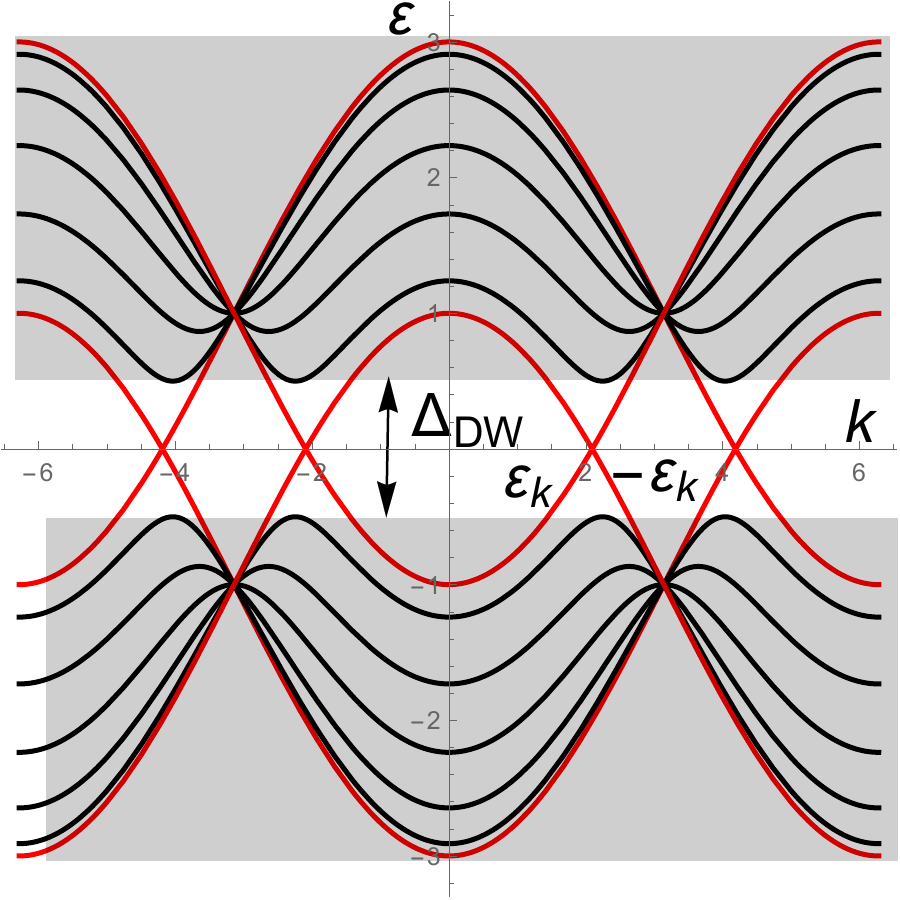}
		\caption{Dispersion laws for ANT of $n = 6$ chains, with doubly-degenerated (black) and non-degenerated (red) modes. The Dirac window of width 
			$\D_{\rm DW} = 1$ is more narrow than for the ZNT in Fig. \ref{figa}}
		\lb{figz}
	\end{figure}
	
	It should be also noted that, unlike a complete similarity between the ZNT and ANR spectra, there is an important difference between those for ANT and 
	ZNR (the latter	having no DLM's at all but presenting instead a special edge mode \cite{Nakada1996}, \cite{PL2022}). 
	
	Then the expansion of local operators (an analog to Eq. \ref{aKK}), reduced to only DLM eigen-operators, results as: 
	\be
	a_{p,j,s} = \frac 1{2\sqrt{nN}}\sum_k {\rm e}^{ikp_s}u_s^\dagger \psi_k,
	\lb{zKK}
	\ee
	with 2-spinors:
	\bea
	&& \psi_k = \left(\begin{array}{c}
		\psi_{0,k;3} \\
		\psi_{0,k;4}
	\end{array}\right),\, u_1 = \left(\begin{array}{c}
		-1 \\
		1
	\end{array}\right) = - u_3,\nn\\
	&&\qquad\qquad\quad u_2 = \left(\begin{array}{c}
		-1 \\
		-1
	\end{array}\right) = -u_4.
	\lb{us}
	\eea
	Due to relative simplicity of expansions by Eq. \ref{zKK} in only 2 DLM's, compared to the ZNT case by Eq. \ref{aKK} with up to 8 DLM's, we begin the next 
	consideration of impurity effects just from the ANT case. 
	
	\section{Impurity effects on ANT}
\begin{figure}[h]
		\centering  
		\includegraphics[width=9cm]{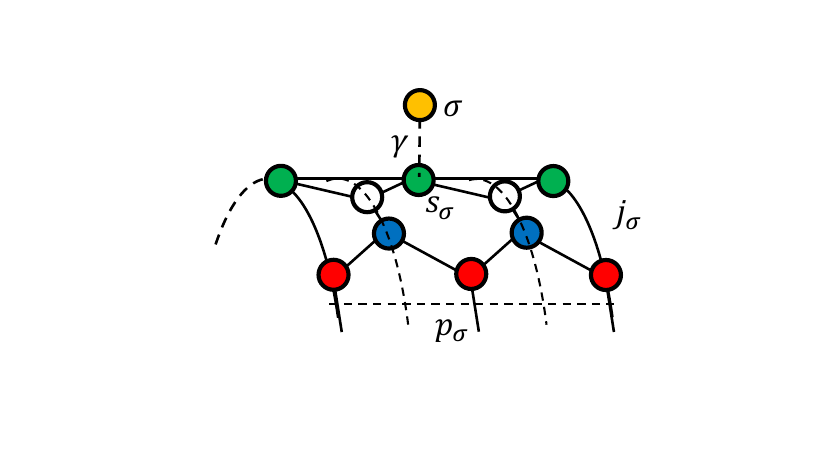}
		\caption{A fragment of ANT with $\s$-th impurity adatom (orange) linked by hybridization $\g$ to its nearest neighbor host atom (green) at the 
			($p_\s,j_\s,s_\s$)-site.}
		\label{fig7}
	\end{figure}

\lb{Imp}
	Now we can consider impurity effects on the above described NT's. The simplest Lifshitz isotopic perturbation model \cite{Lifshitz} is known not to produce 
	impurity resonance effects in NR's, both in ANR and ZNR \cite{PL2022}, therefore we begin from the more effective Anderson hybrid model \cite{Anderson}, 
	presenting its perturbation Hamiltonian for the ANT case (with use of 2-spinors by Eq. \ref{zKK}) as:
	\bea
	H_{\rm AZ} & = & \sum_\s \left[\e_{res}  b_\s^\dagger b_\s\right.\nn\\
	& + &\left. \frac{\g}{2\sqrt{nN}}\sum_k\left({\rm e}^{ikp_\s} b_\s^\dagger u_{s_\s}^\dagger \psi_k  + {\rm h.c.}\right)\right].
	\lb{HA}
	\eea
	It describes impurity adatoms with their resonance energy $\e_{res}$ (laying inside the host DLM range) and corresponding local Fermi operators $b_\s$ at random 
	positions $\s$, linked through the hybridization parameter $\g$ to its nearest neighbor host atom at $s_\s$ site in $p_\s$ segment of $j_\s$ chain (see Fig. 
	\ref{fig7}). The random $p_\s$, $j_\s$, and $s_\s$ values are distributed uniformly with a low overall concentration: $c = (4nN)^{-1}\sum_\s 1 \ll 1$. 
	
	The next consideration goes in terms of (advanced) Green's functions (GF's) whose Fourier-transform in energy:
	\be
	\llang A|B\rrang_\e = \frac i{\pi} \int_{-\infty}^0  {\rm e}^{i(\e - i0)t}\langle\left\{A(t),B(0)\right\}\rangle dt.
	\lb{gf}
	\ee
	includes the grand-canonical statistical average: $\langle O\rangle = {\rm Tr}\,\left[{\rm e}^{-(H - \m)/k_{\mathrm{B}}T}O_H(t)\right]\bigl/\,{\rm Tr}
	\,\left[{\rm e}^{-(H - \m)/k_{\mathrm{B}}T}\right]$ of a Heisenberg operator $O(t) = {\rm e}^{iHt}O{\rm e}^{-iHt}$ under a Hamiltonian $H$ with chemical 
	potential $\m$ and the anticommutator $\{.,.\}$.
	
	As known \cite{Zubarev1960,Economou1979}, GF's satisfy the equation of motion:
	\be
	\e \llang A|B\rrang_\e = \langle\left\{A(0),B(0)\right\}\rangle + \llang [A,H]|B\rrang_\e.
	\lb{de}
	\ee
	In what follows the energy sub-index at GF's is mostly omitted (or enters directly as its argument). 
	
	Consider now the GF $2\times 2$ matrix $\hat G(k,k') = \llang \psi_k|\psi_{k'}^\dagger\rrang$ made of $\psi$-spinors by Eq. \ref{zKK}. In absence of 
	impurities, with use of the Hamiltonian $H$ by Eq. \ref{HZ}, the explicit solution for this GF turns $k$-diagonal: $\hat G(k,k') \to  \d_{k,k'}
	\hat G^{(0)}(k)$, where
	\be
	\hat G_0(k) =\left(\e - \e_k\hat \t_3\right)^{-1}
	\lb{G0}
	\ee
	with the Pauli matrix $\hat\t_3$.
	
	When passing to the disordered system with its Hamiltonian extended to $H + H_{\rm AZ}$, we get the equation of motion for the $k$-diagonal GF matrix, $\hat G(k,k) 
	\equiv \hat G(k)$:
	\be
	\hat G(k) =  \hat G_0(k) + \frac \g{2\sqrt{nN}}\sum_\s {\rm e}^{-ikp_\s} \hat G_0(k)u_{s_\s}\llang b_\s|\psi_k^\dagger\rrang,
	\lb{Gk}
	\ee
	and then its solution is generally sought in the self-energy form:
	\be
	\hat G(k) = \left(\hat G_0^{-1}(k) - \hat \S_k\right)^{-1},
	\lb{se}
	\ee
	including the self-energy matrix $\hat \S_k$. To find it, we continue the chain of equations of motion, now for the mixed (impurity-DLM) row-vector GF:
	\be
	\llang b_\s|\psi_k^\dagger\rrang (\e - \e_{res}) = \frac \g{2\sqrt{nN}}\sum_{k'} {\rm e}^{ik'p_\s}u_{s_\s}^\dagger \hat G(k',k).
	\lb{bps}
	\ee
	This gives the first contribution to $\hat \S_k$ from its term with $k' = k$ used in Eq. \ref{Gk}:
	\be
	\frac {\g^2}{4nN}\sum_\s \frac{u_{s_\s}u_{s_\s}^\dagger}{\e - \e_{res}} = \frac{c\g^2}{\e - \e_{res}}.
	\lb{t0t1}
	\ee
	It is then extended by writing down the equation of motion for the resting terms with $k' \neq k$ in the r.h.s. of Eq. \ref{bps}:
	\bea
	&& \hat G(k',k) = \frac \g{2\sqrt{nN}}\sum_{\s'} {\rm e}^{-ik'p_{\s'}}\nn\\
	&& \qquad\qquad\qquad \times \, \hat G_0(k')u_{s_{\s'}}\llang b_{\s'}|\psi_k^\dagger\rrang,
	\lb{Gkk}
	\eea
	and choosing the term with $\s' = \s$ in its r.h.s. This generates the (scalar) impurity self-energy $\S_0 = \g^2 G_0$ with the DLM locator GF:
	\be
	G_0 = \frac 1{4nN}\sum_k u_{s_\s}^\dagger\hat G_0(k)u_{s_\s} =  \frac 1{4nN}\sum_k {\rm Tr}\hat G_0(k),
	\lb{g0}
	\ee
	which enters the modified factor $(\e - \e_{res} - \S_0)$  in Eq. \ref{bps}. Then the solution for $\hat G(k)$ in the simplest T-matrix approximation for 
	self-energy reads:
	\be
	\hat G(k) = \left[\e - cT(\e)\hat\t_0 - \e_k\hat\t_3\right]^{-1},
	\lb{GT}
	\ee
	with the scalar T-matrix:
	\be
	T(\e) = \frac{\g^2}{\e - \e_{res} - \S_0}.
	\lb{Te}
	\ee
	The next important GF, the DLM locator, is calculated by usual passing from $k$-summation to integration: 
	\be
	G_0(\e) = \frac \e{4n\pi}\int_{0}^{2\pi} \frac {dk}{\e^2 - \e_k^2}.
	\lb{loc}
	\ee
	Its analytic expression (see Appendix) is:
	\bea
	&& G_0(\e) = \frac{i}{4n}\left\{\frac{\theta\left[(1 -\e)(3 + \e)\right]}{\sqrt{(3 + \e)(1 - \e)}}\right.\nn\\
	&& \qquad\qquad\qquad\qquad + \left.\frac{\theta\left[(1 + \e)(3 - \e)\right]}{\sqrt{(3 - \e)(1 + \e)}}\right\},
	\lb{ap}
	\eea
	approximated in the low-energy range as:
	\be
	G_0(\e) \approx \frac i{2n\sqrt 3}\left(1 + \frac{\e^2}{3}\right).
	\lb{app}
	\ee
	
		More generally, the DLM locator is defined as
	\be
	G(\e) = \frac 1{4\pi N} \sum_k {\rm Tr}\, \hat G(k),
	\lb{Gh}
	\ee
	with the diagonal GF matrix $\hat G(k)$ by Eq. \ref{se}. Within the T-matrix approximation by Eq. \ref{GT}, it results simply as:
	\be
	G(\e) = G_0(\tilde\e),
	\lb{Ge}
	\ee
	with $\tilde\e = \e - cT(\e)$ used instead of $\e$ in Eqs. \ref{ap} or \ref{app}. The resulting $G(\e)$ real and imaginary parts as shown in Fig. \ref{fig8} for 
	the choice of Cu impurities with $\e_{res} = 0.03$, $\g = 0.3$ \cite{Irmer2018} and $c = 0.08$ only slightly differ from those for unperturbed $G_0(\e)$ within 
	the $|\e - \e_{res}| \lesssim \g^2|G_0|$ range.
	\begin{figure}[h]
		\centering
		\includegraphics[width=7cm]{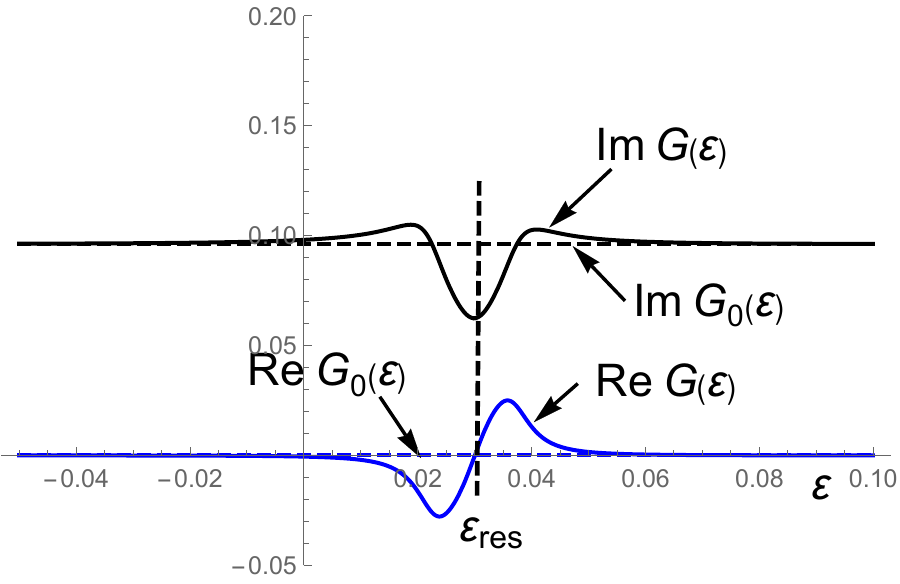}
		\caption{Real (blue) and imaginary (black) parts of the locator function $G(\e)$ compared to those by the unperturbed $G_0(\e)$ (dashed) for ANT with $n = 12$ 
			at Cu impurity concentration $c = 0.08$.}
		\lb{fig8}
	\end{figure}
	
	Another set of elementary excitations in the disordered system, that due to impurity atoms, defines the impurity locator GF: $G_{imp} = N^{-1}\sum_\s \llang 
	b_\s|b_\s^\dagger\rrang$, and its solution in the same approximation reads:
	\be
	G_{imp}(\e) = cT(\e)/\g^2.
	\lb{bb}
	\ee
	Together, the DLM and impurity locators, Eqs. \ref{Ge}, \ref{bb}, define the low-energy density of states (DOS) as $\r(\e) = \r_h(\e) + \r_{imp}(\e)$ with its host 
	and impurity parts: 
	\be
	\r_h(\e) = \frac 2\pi \,{\rm Im}\,G(\e),\quad \r_{imp}(\e) =  \frac 2\pi\,{\rm Im}\,G_{imp}(\e)
	\lb{re}
	\ee
	(taking the account of 2 spin values). 
	\begin{figure}[h]
		\centering
		\includegraphics[width=7cm]{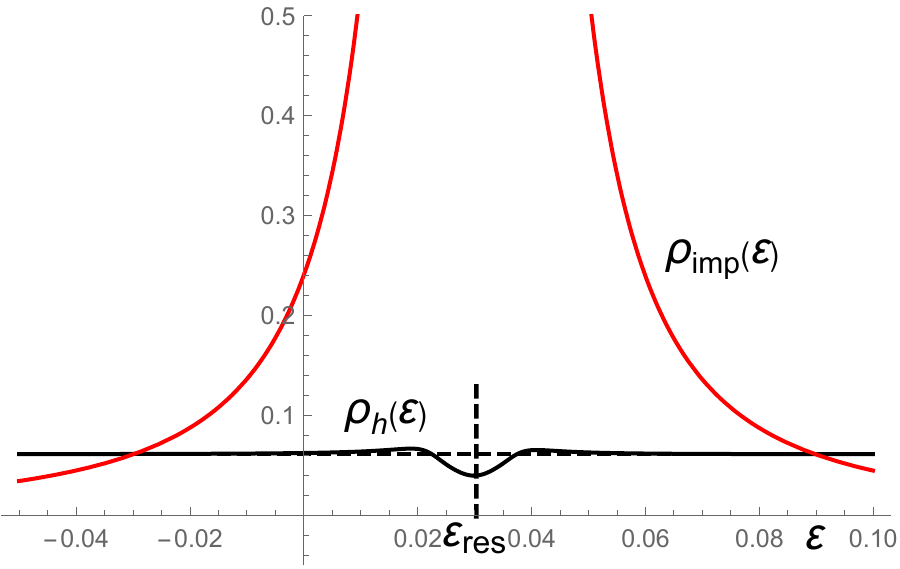}
		\caption{DOS parts near the impurity resonance level for the same system as in Fig. \ref{fig8}.}
		\lb{rho}
	\end{figure}
	
	In a disordered ANT, host DLM's contribute with $1/n$ charge carriers per site and impurities do with $c$ carriers per site, so defining the Fermi level $\e_{\rm F}$ 
	from the equation:
	\be
	\frac 1n + c = \int_{-3}^{\e_{\rm F}}\r(\e)d\e
	\lb{eF}
	\ee
	(integrated from the bottom of DLM range). Then, in the simplest approximation of Im$G(\e) \approx {\rm Im} G_0(\e) \approx 2/(\sqrt 3\pi n)$  (dashed line in Fig. 
	\ref{rho}), the Fermi level dependence on impurity concentration $c$ results:
	\be
	\e_{\rm F}(c) \approx \frac{c}{c + c_{\ast}}\,\e_{res},
	\lb{eFc}
	\ee
	where $c_{\ast} = [\g G_0(\e_{res})]^2$. Its fast initial growth: $\e_{\rm F}(c) \approx (c/c_{\ast})\e_{res}$ at $c \lesssim c_{\ast}$, changes to a slow approach 
	of $\e_{res}$: $\e_{\rm F}(c) \approx \e_{res} - (c_{\ast}/c)\e_{res}$ at $c \gtrsim c_{\ast}$ (see Fig. \ref{eFc}).
	\begin{figure}[h]
		\centering
		\includegraphics[width=7cm]{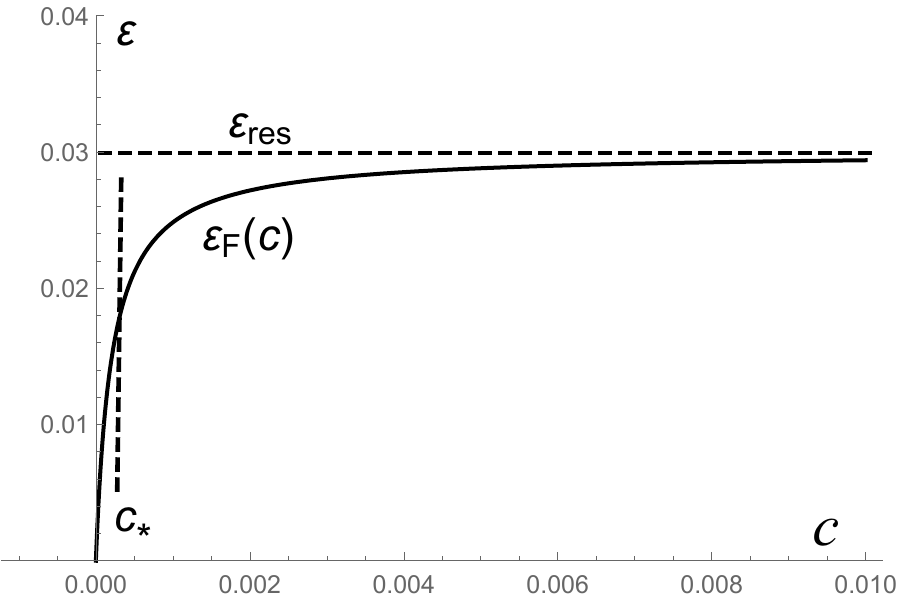}
		\caption{Fermi level in function of impurity concentration for the system by Figs. \ref{fig8}, \ref{rho}.}
		\lb{eFc}
	\end{figure}
	
	The next analysis of low energy spectra in this disordered system follows the lines of similar cases by Refs. \cite{ ILP1987,Pogorelov2020, PL2022}. Thus, the modified 
	dispersion laws are obtained from the standard equation \cite{Bonch}:
	\be
	{\rm Re} \det \hat G^{-1}(k) = 0,
	\lb{disp}
	\ee
	which splits in two scalar equations:
	\be
	\e - c\,{\rm Re}\,T(\e) = \pm \e_k.
	\lb{disp1}
	\ee
	They appear as cubic equations for energy in function of momentum, $\e(k)$, and their analytic solutions, though standard, are rather cumbersome. But they can be 
	greatly simplified within the relevant low energy range by using the linearized dispersion law:
	\be
	\e_k \approx \frac{\sqrt 3}2 k
	\lb{dl}
	\ee
    with momentum $k$ referred to the Dirac point and Fermi velocity $\sqrt 3/2$, and solving Eq. \ref{disp1} for this momentum in function of energy \cite{PL2022}:
	\be
	\pm k_\e = \pm \,2\, \frac {\e - c\,{\rm Re}\,T(\e)}{\sqrt 3}.
	\lb{ke}
	\ee
	An example of such solutions for ANT with $n = 12$ and $c = 0.08$ in Fig. \ref{displ} demonstrates how coupling of each $\pm \e_k$ mode to the impurity $\e_{res}$ 
	mode forms resonance splitting of hybridized $\pm \e_{1k}$ and $\pm \e_{2k}$ modes at $k = 0$ to the interval of $\approx \sqrt{\e_{res}^2 + 4c\g^2}$ around $\e_{res}/2$.
	\begin{figure}[h]
		\centering
		\includegraphics[width=8cm,angle=90]{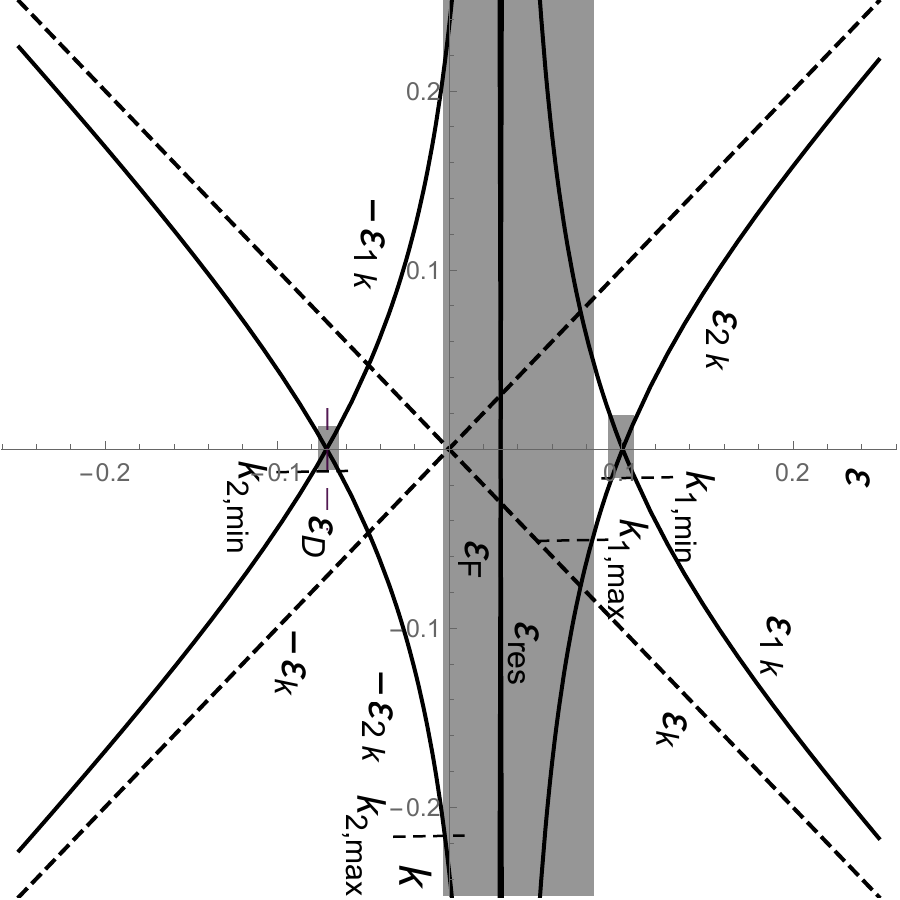}
		\caption{Modified dispersion laws $\e_{1,k}$ and $\e_{2,k}$ by Eq. \ref{ke} for for the system as in Figs. \ref{fig8}-\ref{eFc} compared to the 
			unperturbed $\e_k$ and to the impurity level $\e_{res}$ (the momentum $k$ being referred to the $K$-point). The mobility gaps (shadowed), define 
			the mobility edges $k_{1,min}$, $k_{2,min}$, and $k_{2,max}$ for quasiparticle momenta.}
		\lb{displ}
	\end{figure}
	
	Next, this $k_\e$-form is used in the important test of dispersion law validity for a disordered system, the Ioffe-Regel-Mott (IRM) criterion \cite{IoffeRegel, 
		Mott}:
	\be
	k_\e v_\e \gtrsim \t^{-1}(\e),
	\lb{IRM}
	\ee
	with the quasiparticle group velocity $v_\e = (\partial k_\e/\partial \e)^{-1}$ and its inverse lifetime $\t_\e^{-1} = c\,{\rm Im}\,T(\e)$, that is the 
	quasiparticle mean free path to be longer of its wavelength. 
	
	Each $\e$ value that converts $\gtrsim$ into $\approx$ in Eq. \ref{IRM} gives an estimate for a mobility edge $\e_{mob}$, separating the ranges of band-like and 
	localized states in the spectrum. An important rule for these states in a multi-mode system is that they can not coexist, that is, if, for a certain energy, the 
	IRM criterion does not hold for at least one mode, all other modes at this energy should be also localized \cite{Mott}. 
	
	With use of Eqs. \ref{ke}, \ref{Te}, such equation for mobility edges can be written explicitly in the form:
	\be
	\frac{c\g^2\G}{D^2(\e )}\left|1 + c\g^2\frac{D^2(\e ) - 2\G^2}{D^4(\e )}\right| \approx \left|\e - c\g^2\frac{\e - \e_{res}}{D^2(\e )}\right| 
	\lb{IRM1}
	\ee
	\vspace{3pt}
	with $D^2(\e ) = (\e - \e_{res})^2 + \G^2$ and $\G = \g^2 G_0(\e)$. Then, using the $G_0(\e)$ value by Eq. \ref{ap}, this equation can be solved numerically to 
	estimate all possible $\e_{mob}$ values and so delimit the band-like and localized energy ranges in ZNT with impurities at given disorder parameters ($\e_{res}$, $\g$, 
	$c$) and of NT structure ($n$) as shown in Fig. \ref{displ}. 
	
	Here one localized range is found at the lower limit of resonance splitting, near the shifted Dirac energy $-\e_s$ at all $c > 0$, being of width $\approx c\g^2/\e_{res}$. 
	Another localized range emerges above it, around $\e_{res}$, when $c$ reaches a certain critical value $c_0$. And at yet higher critical concentration, $c_1 \gg c_0$, the 
	latter range gets split in two, due to a specific interplay (when going away from the Dirac point) between the growing momentum $k_\e$, decreasing group velocity $v_k$, and 
	increasing inverse lifetime $\t_\e^{-1}$  of hybridized modes. A more detailed description of these restructured spectra for different nanostructures follows below. 
	
	\section{Impurity effects on ZNT}\lb{ZNT}
	
	It is also of interest to extend the above approach to another NT topology, namely, to the more involved ZNT case. To simplify description of low energy impurity 
	resonances here, we again restrict the expansions of local operators by Eqs. \ref{apj} and \ref{psi} in 4-spinors $\psi(k,q)$ with unitary matrices $\hat U(k,q)$, to
	 only DLM's $q = K,K'$. Then the ZNT perturbation Hamiltonian results, instead of Eq. \ref{HA} for ZNT, in the form: 
	\bea
	&& H_{\rm Z} = \sum_\s\e_{res}b_\s^\dagger b_\s + \frac{\g}{2\sqrt{nN}} \sum_{k,\s} \left\{{\rm e}^{ikp_\s}b_\s^\dagger\right.\nn\\
	&&  \times \left[{\rm e}^{iKj_\s}[u^\dagger(k,K;\s)\psi(k,K)]_{j_\s}\right.\nn\\
	&& \left.\left. +\, {\rm e}^{iK'j_\s}[u^\dagger(k,K';\s)\psi(k,K')]_{j_\s}\right] + {\rm h.c.}\right\},
	\lb{AMA}
	\eea
	where the row spinor $u^\dagger(k,q;\s)$ is just the $j_\s$-th row of $\hat U^\dagger(k,q)$.
	
	Next we consider the 4$\times$4 GF matrices $\hat G(k,q;k',q') \equiv \llang  \psi_{k,q}|\psi_{k',q'}^\dagger\rrang$ and the related equation of motion with the 
	Hamiltonian $H + H_{\rm Z}$ for the choice of $K$-modes GF:
	\bea
	&&\hat G(k,K;k',K) = \d_{k,k'}\hat G^{(0)}(k,K)\nn\\
	&&\qquad + \frac{\g}{2\sqrt{nN}}\sum_\s {\rm e}^{-i(kp_\s + Kj_\s)}\hat G_0(k,K)\nn\\
	&&\qquad\qquad\qquad \times\,u(k,K;\s)\llang b_\s|\psi_{k',K}^\dagger\rrang, 
	\eea
	where $G_{f,f'}^{(0)}(k,K) = \d_{f,f'}(\e - \e_{k,K,f})^{-1}$ and the column spinor $u(k,q;\s)$ is the $j_\s$-th column of $\hat U(k,q)$.
	
	Then the equation (similar to Eq. \ref{bps} for the ZNT case) for the mixed  GF, $\llang b_\s|\psi_{k',K}^\dagger\rrang$:
	\bea
	&&\llang b_\s|\psi_{k',K}^\dagger\rrang\left(\e - \e_{res}\right) = \frac{\g}{2\sqrt{nN}}\sum_{\k''} {\rm e}^{i(k''p_\s + Kj_\s)}\nn\\
	&& \qquad\qquad\qquad\qquad\times u^\dagger(k'',K;\s)\hat G(k'',k')
	\eea
	leads (in the same way as to Eq. \ref{GT}) to the T-matrix solution for momentum-diagonal GF matrix $\hat G(k,K) \equiv \hat G(k,K;k,K)$:
	\be
	\hat G(k,K) = \left\{\left[\hat G^{(0)}(k,K)\right]^{-1} - cT(\e)\right\}^{-1}
	\ee
	where the T-function for this case differs from that by Eq. \ref{Te} only by the form of its locator $G_0(\e)$. Using Eq. \ref{KKa}, it results here as:
	\be
	G_0(\e) = \frac{4i}{n\sqrt{1 - (\e/2)^2}}.
	\lb{loca}
	\ee
	Then comparison with Eq. \ref{ap} shows that the impurity level damping for this case turns $\approx 8\sqrt 3$ times stronger than for ANT at the same $n$ number, mostly due to the above indicated greater relative weight of DLM's in ZNT than in ANT spectra. This produces strongly different behaviors of IRM mobility edges and qualitatively different structures of localized and band-like spectra in these two nanosystems.

	\section{Comparison with other carbon nanosystems}\lb{Comp}
	
		\begin{figure}[!h]
		\centering
		\includegraphics[width=8cm]{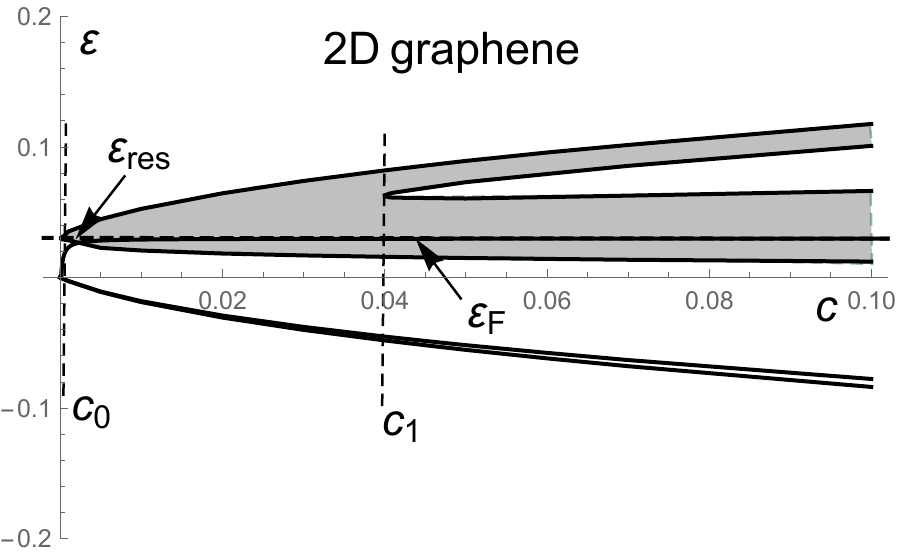}
		\includegraphics[width=8cm]{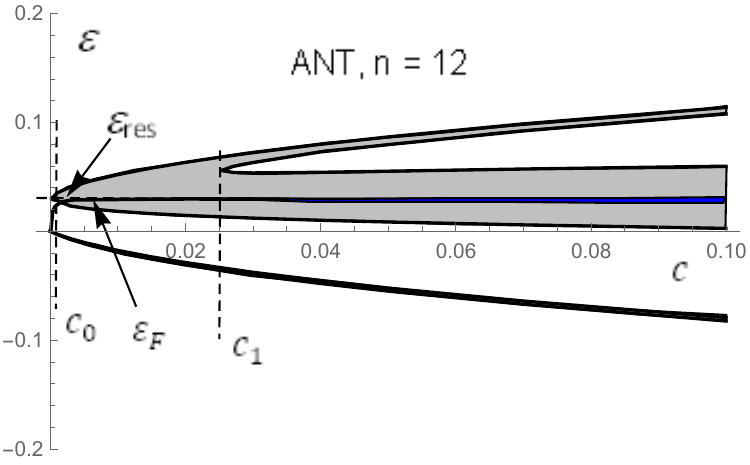}
				\includegraphics[width=8cm]{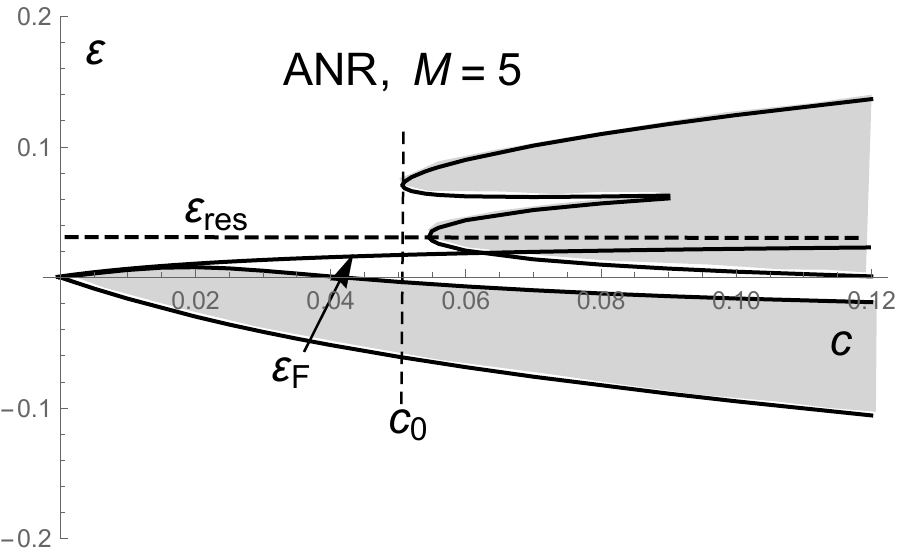}
		\includegraphics[width=8cm]{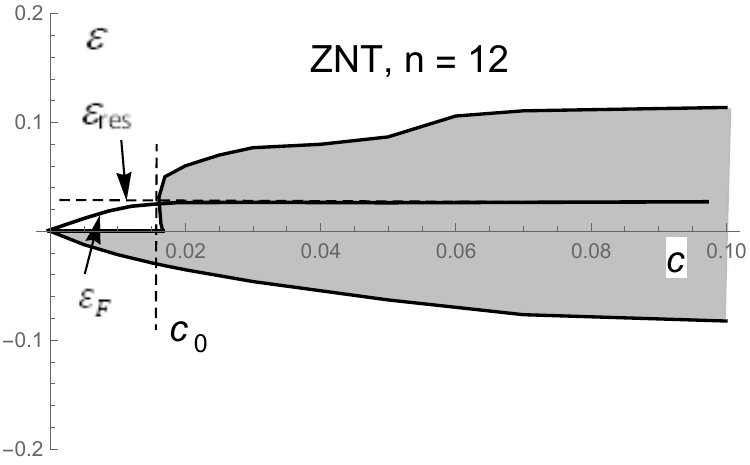}
		\caption{Development of localized ranges (shadowed) in quasiparticle spectra of: 2D graphene, armchair nanotube (with a narrow blue range of 
			divergence for GF group expansion, see Sec. \ref{Beyond}),  armchair nanoribbon, and zigzag nanotube with growing Cu impurity concentration.}
		\label{4figs} 
	\end{figure}  
The low-energy spectrum restructuring under impurity disorder effect is suitably illustrated by a diagram of mobility edges $\e_{mob}$ between the localized and 
band-like energy ranges in function of impurity concentration $c$. Such diagrams in Fig. \ref{4figs} permit to compare the effects of Cu impurities on Dirac modes 
in the previously considered 2D graphene \cite{Pogorelov2020} and ANRs \cite{PL2022} together with the above obtained results for ZNT and ANT. 

Some general features, noted for the ANT case, are observed in all of them:
	i) formation of a localized range (mobility gap) around the resonance level $\e_{res}$, at reaching a certain critical concentration $c_0$, ,
	
	ii) presence of another localized range around the shifted down Dirac level $\e_D$, being mostly narrower but existing at all $c > 0$, 
	
	iii) opening, at a certain higher critical concentration $c_1 \gg c_0$, of a narrow band range within the $\e_{res}$-related mobility gap. 
	
	But this comparison also reveals notably different sensitivity of the corresponding DLM's to the impurity resonance level, depending both on their topological 
	properties (absence or presence of edges and the edge types) and on discrete transversal numbers of chains in a system. Within the IRM formalism, for given 
	impurity parameters $\e_{res}$ and $\g$, it depends on the host system through its locator function $G_0(\e)$, like those by Eqs. \ref{ap}, \ref{loca}. This can be 
	further compared with the previously found $G_0(\e)$ values for 2D graphene: $\e/\sqrt 3$ \cite{YDV2021} and for ANR with $M$ carbon chains: $4/(M + 1)$ \cite{PL2022}.
	
	Thus, the impurity-induced localization first occurs at an energy very close to $\e_{res}$ and the related critical concentration $c_0$ can be estimated from 
	Eq. \ref{IRM1} by setting $\e = \e_{res}$ there. It results generally in: 
	\be
	c_0 = \frac{[\g G_0(\e_{res})]^2}2\left[1 + \sqrt{1 + 4\frac{\e_{res}}{\g^2  G_0(\e_{res})}}\right],
	\lb{c0}
	\ee
	and for the instance of ANT with $n = 12$ it gives $c_0 \approx 1.7\cdot 10^{-4}$ in a reasonable agreement with the numerical calculation result shown in Fig. \ref{4figs}. 
	With further growth of $c > c_0$, a continuous range of localized states (mobility gap) appears  around $\e_{res}$, of width growing as $\D_{mob} \sim \g\sqrt{c - c_0}$. 
	
\begin{figure}[h]
		\centering
		\includegraphics[width=6cm]{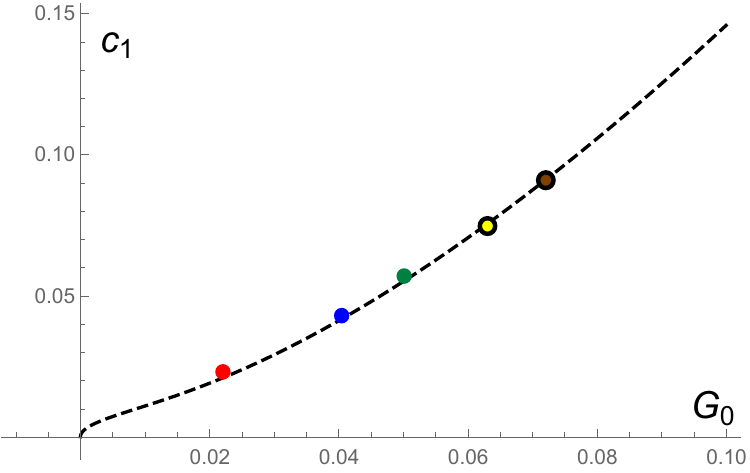}
		\caption{Upper critical concentration $c_1$ for Cu impurities in: ZNT with 
			$n = 50$ (brown), ANR with $M = 47$ (yellow), ANT with $n = 6$ (green), 2D graphene (blue), and ANT with $n = 12$ (red), 
			together with the fitting curve by Eq. \ref{c1} in function of locator $G_0$ value.}
		\label{critvsG}
	\end{figure}

	Then, at reaching another critical concentration $c_1 \gg c_0$, a certain window of band-like states opens inside the mobility gap, due to the before discussed faster 
	resonance splitting between the initial $\e_k$ and $\e_{res}$ modes than these split modes damping. This $c_1$ value is also estimated from the numerical solution of 
	IRM Eq. \ref{IRM}. It can be presented in function of the single $G_0$ parameter as shown in Fig. \ref{critvsG}. Strictly speaking, this is only possible for 1D nanosystems 
	where the low-energy locator $G_0(\e)$ is practically constant, defined by their topology and discrete width numbers. This dependence can be reasonably fitted by the 
	formula:
	\be
	c_1 \approx \sqrt{0.01 G_0 + 20 G_0^3}.
	\lb{c1}
	\ee
	
	The IRM test also indicates a similar mobility window to open under the same impurities in 2D graphene with linear $G_0(\e)$ behavior. The resulting $c_1$ value 
	qualitatively agrees with the approximation by Eq. \ref{c1} at the choice of $G_0 = G_0(\e_{op})$, $\e_{op}$ being just the energy where the mobility window first 
	opens (as included into Fig. \ref{c1vsG}).
	
	Notably, the $G_0$ parameter decreases with the nanotube width as $\sim 1/n$, producing respective decrease of $c_1$ and so making the system spectrum more 
	sensible to impurity resonances. Thus, the $c_1$ value for ZNT with $n > 8$ should turn already below of that for 2D graphene (despite the latter could be 
	formally thought as the $n \to \infty$ limit), underlying the importance of topological factors in these effects.
	
	However, as it was already noted above, such widening of a nanotube would produce a similar narrowing of the Dirac window $\D_{DW}$ in its spectrum, delimiting 
	the range of possible impurity effects. Therefore, the optimal conditions for them should be sought from a certain compromise between the parameters of impurity 
	(energy level $\e_{res}$, hybridization $\g$, and concentration $c$) and of host NT (topological type and width $n$). Thus, for the considered Cu impurities, we 
	estimate admissible width limits for ANT: $n \lesssim 40$, ZNT: $n \lesssim 60$, and ANR: $M \lesssim 35$. Their comparison with the $c_1$ estimates in Fig. 
	\ref{critvsG} suggests possibility for the narrow conductivity window above $\e_{res}$ in ANT, graphene and maybe in ZNT, but hardly in ANR (though the latter can 
	provide a similar window below $\e_{res}$).

		\section{Twisted nanotubes}\lb{TNT}
		
		Yet more general structure of a TNT is intermediate between the above considered ANT and ZNT, with its unit cell being defined by two natural numbers 
		$n$ and $m$, based on the chiral vector ${\bold{C}_{n,m}} = n{\bold{a}}_1 + m{\bold{a}}_2$ and its orthogonal longitudinal vector ${\bold{T}_{n,m}} 
		= [(2m + n){\bold{a}}_1 - (2n + m){\bold{a}}_2]/R_{n,m}$ (where $R_{n,m}$ is the greatest common divisor of $2m + n$ and $2n + m$). There are altogether 
		$N_{n,m} = 4C_{n,m}T_{n,m}/(\sqrt 3 a^2)$ atomic positions in this cell, as shown for the example of $n = 4, m = 1$ in Fig. \ref{twist}a. 
		
		TNT structure differs qualitatively from the limiting ANT and ZNT ones in that it has a single period along ${\bold C}_{n,m}$ but repeated periods along the 
		longitudinal ${\bold T}_{n,m}$, defining purely 1D translational symmetry. The resulting spectrum consists of $N_{n,m}$ purely 1D modes and it contains DLM's under 
		the condition of $n - m = 3l$ with a natural $l$ \cite{Kane} (which passes to ANT at $l = 0$ and ZNT at $m = 0$). Then multiple 1D Brillouin zones (BZ) in such a 
		NT with their longitudinal period $\tilde{T}_{n,m} = 2\pi/T_{n,m}$ result just commensurable with the Dirac points in multiple 2D BZ of planar graphene. An example 
		of TNT with $n = 4, m = 1$ in Fig. \ref{twist}b shows such matching of its 1D BZ's to some of graphene Dirac points. 
		\begin{figure}[h]
			\centering
			\includegraphics[width=9cm]{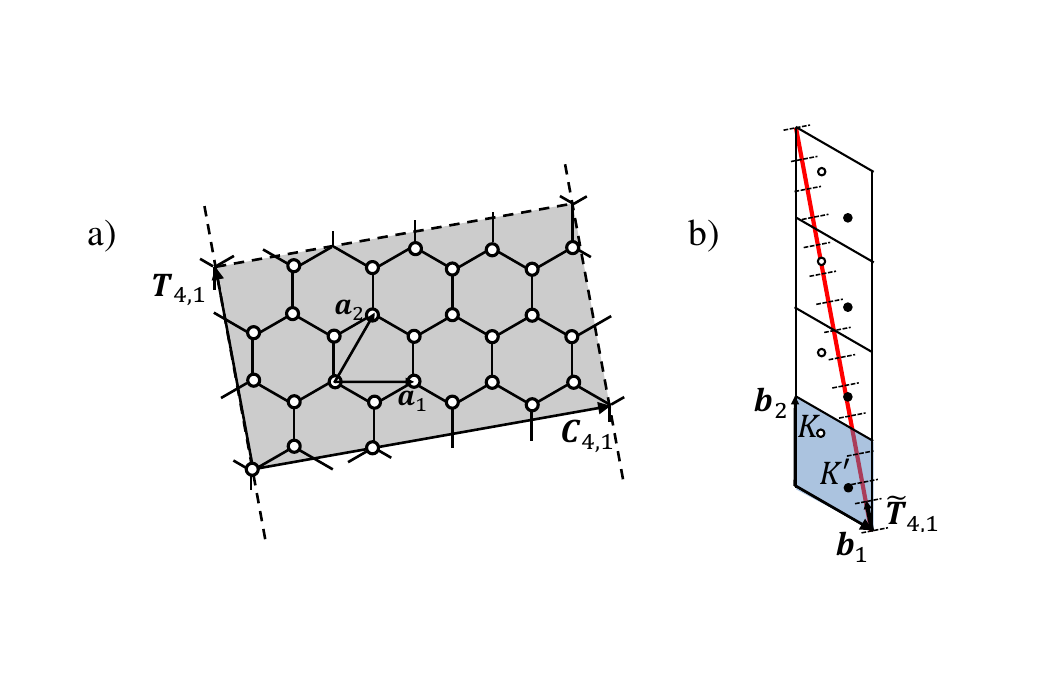}
			\caption{a) The unit cell of $n = 4, m = 1$ TNT (unfolded) with its chiral ${\bold C}_{4,1}$ and longitudinal ${\bold T}_{4,1}$ base vectors, containing $N_{4,1} 
				= 28$ atomic positions. b) The sequence of 14 1D Brilloin zones by this TNT (red line) along the base vector ${\tilde{\bold T}_{4,1}} = 2\pi{\bold T}_{4,1}/T^2_{4,1}$, 
				it matches to certain Dirac points from multiples of 2D graphene BZ (shadowed) with its base vectors ${\bold b}_{1,2}$, exactly at $2/3$ or $1/3$ of the $\tilde 
				T_{4,1}$ periods (dotted segments).}
			\lb{twist}
		\end{figure} 
		
		A treatment of impurity effects on TNT can be done within the above restriction to only DLM's with its results mostly defined by the related value of locator $G_0$. 
		But here Eq. \ref{loc} should be modified by changing the $4n$ factor to (possibly much bigger) $N_{n,m}$ and also the Fermi velocity $\sqrt 3/2$ in Eq. \ref{dl} to 
		a much higher $T_{n,m}\sqrt{3}/2$, resulting in much lower $G_0$ values. Then, accordingly to the results by Sec. \ref{Comp}, much lower critical impurity concentrations 
		and much higher sensitivity of (properly chosen) TNT to impurity effects can be expected. A more detailed discussion of these issues will be given elsewhere.

	\section{Beyond T-matrix approximation}\lb{Beyond}
	
	Besides the most common approach to spectra of disordered systems through the single-impurity scattering in terms of T-matrix, there are its certain extensions. One of 
	them uses the self-consistent approximation to this T-matrix \cite{Freed}, another is based on group expansions of self-energy \cite{ILP1987, LP2015} in series of terms 
	corresponding to wave scatterings by various clusters of increasing number of impurities. 
	
	\subsection{Self-consistent approximation}\lb{selfcons}
	Let us begin from the self-consistent approximation where the T-matrix is written as:
	\be
	T_{s-c}(\e) = \frac{\g^2}{\e - \e_{res} - \g^2 G_{s-c}(\e)},
	\lb{Tsc}
	\ee
	with the self-consistent locator $G_{s-c}(\e) = G_0[\e - c T_{s-c}(\e)]$.
	
	Then, using the above approximated expression by Eq. \ref{app}, we obtain the self-consistency equation for $G_{s-c}(\e)$:
	\bea
	&& i \sqrt 3\,n\,G_{s-c}(\e) + 1\nn\\
	&& \qquad + \frac 13 \left[\e - \frac{c\g^2}{\e - \e_{res} -  \g^2 G_{s-c}(\e)}\right]^2 = 0.
	\eea
	\begin{figure}[h]
		\centering
		\includegraphics[width=8cm]{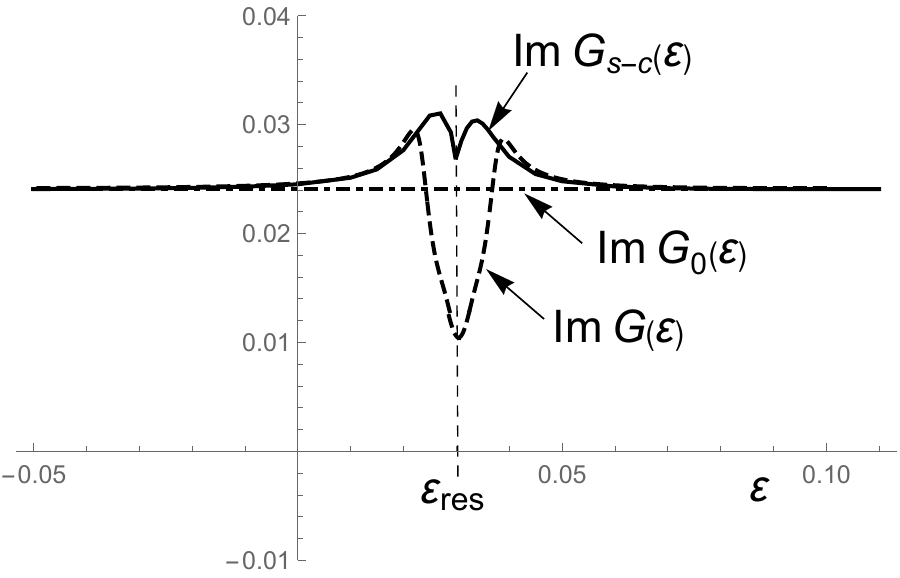}
		\caption{Imaginary parts of locator functions: self-consistent $G_{s-c}(\e)$ (solid), simple T-matrix $G(\e)$ (dashed), and unperturbed $G_0(\e)$ (dash-dotted) for 
			the ANT system as in Figs. \ref{fig8}-\ref{displ}, mostly differing near the resonance level $\e_{res}$.}
		\label{selfcons}
	\end{figure}
	\begin{figure}[h]
		\centering
		\includegraphics[width=8cm]{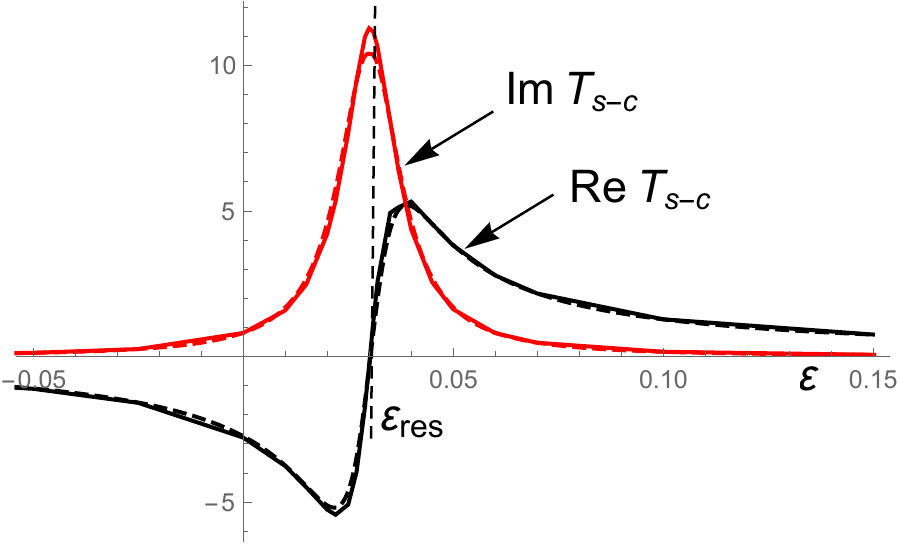}
		\caption{Practical coincidence of self-consistent (solid lines) and simple (dashed lines) T-matrix functions for the same system as shown in Fig. \ref{selfcons}.}
		\label{Tsc}
	\end{figure}
	Its numerical solution for the characteristic case of ANT with $n = 6$ and $c = 0.015$ provides the real and imaginary parts by $G_{s-c}(\e)$ as shown in Fig. 
	\ref{selfcons} in comparison with the same parts of the simple $G_0(\e)$, Eq. \ref{Te}. It demonstrates that the self-consistency correction only slightly changes 
	$G_0(\e)$ in a vicinity of $\e_{res}$, and also such parts of $T_{s-c}$ and $T_0$ are almost coincident (Fig. \ref{Tsc}).  So this change has practically no effect 
	on the IRM results obtained above with use of the simple $G_0(\e)$. So the corresponding mobility diagrams as in Fig. \ref{4figs} remain also valid in the self-consistent 
	approximation.
	
	\subsection{Group expansion}\lb{GE}
	Next, we look for a group expansion (GE) of the self-energy matrix $\hat\S_k(\e)$ in the form:
	\be
	\hat\S_k(\e) = c\hat T(\e)\bigl[\hat 1 + c \hat B_k(\e) + \dots\bigr],
	\lb{GE}
	\ee
	where the sum:
	\be
	\hat B_k(\e) = \sum_{r} \bigl[{\rm e}^{-ikr}\hat A_r(\e) + \hat A_r^2(\e)\bigr]\bigl[\hat 1 - \hat A_r^2(\e)\bigr]^{-1}
	\lb{Bk}
	\ee
	describes the effects of multiple scatterings between pairs of impurities at longitudinal distance $r$ between them through the related scattering matrix 
	$\hat A_r(\e) = \hat T(\e)\hat G_r(\e)$ with the correlator matrix
	\[\hat G_r(\e) = \frac 1{4nN} \sum_{k' \neq k} {\rm e}^{ik'r}\hat G_0(k).\] 
	The omitted terms in the r.h.s. of Eq. \ref{GE} correspond to contributions by clusters of three and more impurities.
	
	Notably, the important specifics of the disordered carbon NT's (and NR's), unlike the commonly studied disordered 3D or 2D crystals, consists in that:
	
	$\ast$ here the longitudinal distance $r$ between different impurities takes only discrete values, namely, $2r$ takes integer values (so the sum $\S_r$ 
	can be done without usual passing to integral $\int dr$) and
	
	$\ast$ this distance can be also {\it zero}. 
	
	Thus, it can be seen from Figs. \ref{fig6}, \ref{fig7} that for any impurity position $\s$, there are $2n - 1$ other positions $\s'$ with the same longitudinal 
	coordinate $p_\s = p_{\s'}$. Such impurity pairs at zero longitudinal distance contribute into $\hat B_k$ with 
	\be
	\hat B_0 = c\left(1 - \frac 1n\right)\bigl[\hat A_0(\e) + \hat A_0^2(\e)\bigr]
	\bigl[\hat 1 - \hat A_0^2(\e)\bigr]^{-1}
	\lb{B0}
	\ee
	and this contribution results dominant over the resting sum $\sum_{r \neq 0}$ in Eq. \ref{Bk} (see in Appendix). 
	
	Consider it in more detail for the example of ANT where all the matrices in the self-energy $\hat \S_k$ can be substituted by scalars: $A_0(\e) = T(\e)G_0(\e)$, 
	and $T(\e)$ can be taken in the form by Eq. \ref{Te}. Then the $B_0(\e)$ contribution to GE by Eq. \ref{GE} is estimated using the explicit form of $A(\e) = 
	\G/(e - \e_{res} - i\G)$ with $\G = \g^2/(2n\sqrt 3)$ to give:
	\be
	B_0(\e) = (1 - \frac 1{2n})\G \frac{\e - \e_{res} - i\G}{(e - \e_{res})^2}.
	\lb{B0} 
	\ee
	For comparison, the lowest degree resting term in $B_k(\e)$ is evaluated with use of the approximation for $A_r(\e)$ by Eq. \ref{Gapp}, as: 
	\be
	\sum_{r \neq 0} {\rm e}^{-ikr} A_r(\e) \approx \frac{i\G\e}{4n\pi^2}\, {\rm Li}_2\left(-{\rm e}^{-ik/2}\right),
	\lb{Bk0}
	\ee
	where the polylogarithmic function \cite{AbS} at $k \ll 1$ is close to ${\rm Li}_2(-1) = -\pi^2/12$. Thus the magnitude of Eq. \ref{Bk0} term turns more than 4 
	orders below of that by Eq. \ref{B0} within the whole low-energy range, while the next terms in $B_k(\e)$ result yet much smaller. 
	
	Then, taking the GE convergence criterion as $c|B_0(\e)| < 1$, the T-matrix validity condition results from Eq. \ref{B0} well approximated by:
	\be
	|\e - \e_{res}| \gtrsim \G\sqrt c,
	\lb{GEc}
	\ee
	and it can only fail in a very narrow vicinity of $\e_{res}$ (the blue range in Fig. \ref{4figs} for ZNT) deeply within the localized range, just confirming 
	localization of states there. In a similar way, this conclusion can be reached for other nanosystems considered here, justifying the above obtained pictures 
	of spectrum restructuring in them.

	\section{Discussion of results}\lb{Disc}
	
	The above obtained results on restructured low-energy quasiparticle spectra in carbon nanosystems can be discussed in the context of disordered 1D crystalline 
	systems generally known not to contain conducting states at any degree of disorder \cite{Anderson1958, Hjort, Delande, Vosk, YaqiTao}. But their presence in the 
	disordered NT's and NR's indicates again the principal qualitative difference of these structures from the strictly 1D chains. Besides the above noted possibility 
	for zero longitudinal distance between different impurity positions, it can be yet illustrated by the behaviors of correlator functions with growing inter-impurity 
	distance $r$: converging as $\sim 1/r^2$ by Eq. \ref{Gapp} for NT's and diverging as $\sim 1/r$ for really 1D chains (as that by single $g_r(\e)$ by Eq. \ref{HG}), 
	making the related GE divergent at all energies. So we can conclude that it is just the presence of additional transversal degrees of freedom in quasi-1D systems 
	that enables their conductivity under disorder \cite{Ando1998, Ando2000, Biel2008}	.
	
	The found intermittence of conducting and localized energy ranges in the considered nanosystems can be then used for their various practical applications. Thus, 
	the most straightforward effects are expected in frequency $\o$- and temperature $T$-dependent electric conductivity, following from the general Kubo-Greenwood 
	formula \cite{Kubo}, \cite{Greenwood} presented here in the form:
	\bea 
	&&\s(\o,T) = \frac{e^2}{\pi}\int d\e \,\frac{f(\e,T) - f(\e',T)}{\o}\nn\\
	&& \times \int_{cond} dk \,v_k(\e)v_k(\e'){\rm Im}G_k(\e)\,{\rm Im}G_k(\e').
	\lb{Kubo}
	\eea
	It includes the Fermi function $f(\e,T) = [{\rm e}^{(\e - \e_F)/T} + 1]^{-1}$, the group velocity $v_k(\e) = [\partial k_\e/\partial\e]^{-1}$, the a.c. shifted energy 
	$\e' = \e + \hbar\o$, and the integration $\int_{cond}$ avoids localized ranges (as those shadowed in Fig. \ref{displ}). 
	
	First of all, consider the simplest d.c. limit:
	\[\lim_{\o \to 0} \frac{f(\e,T) - f(\e',T)}{\o} \to \frac 1{4T{\rm Cosh}^2[(\e - \e_F)/2T]},\]
	which then goes to $\d(\e - \e_F)$ at $T \to 0$, defining
	\[\s(0,0) = \frac{e^2}{\pi}\int_{cond} dk\, v_k^2(\e_F)[{\rm Im}G_k(\e_F)]^2,\]
	and this turns zero for $\e_{\rm F}$ laying within a localized range. But such insulating state can be converted to conducting by applying 	quite a small external gate 
	voltage $V_g$. Thus, for the ANT case of Fig. \ref{displ}, the initial $\e_{\rm F} \approx \e_{res} = 0.03$ (in units of $t \approx 2.8$ eV) could reach the nearest 
	mobility edges with gating either $\approx 100$ meV upwards or $\approx 80$ meV downwards, and the resulting reversible insulator-metal transitions should stay well 
	resolved up to room temperatures. 
	
	Otherwise, a quite sharp threshold in optical conductivity can be reached by applying IR radiation of $\sim 10$ THz (which may be also combined with a slight gate 
	tuning $V_g \sim 5$ meV). A more detailed description of the $\s(\o,T)$ behavior readily follows from the above given T-matrix solutions for $v_k(\e)$ and ${\rm Im}G_k(\e)$. 
	All these effects are most diversified with formation of multiple mobility edges (above the second critical concentration $c_1$). 
	
	The above quantitative results were delimited to a single choice of Cu impurity in its top-position over a host carbon atom, but they can be readily extended to 
	other impurities in different positions, providing a variety of possible values for the relevant $\e_{res}$ and $\g$ parameters and so a much broader field of 
	resulting electronic dynamics. Nevertheless, their qualitative features indicated in the present study should stay proper for all of them.
	
	Yet another practical condition for validity of the above conclusions consists in that a NT (or a NR) should be long enough compared to the localization length $l_{loc}$ 
	of quasiparticle states near the mobility edges. The latter can be estimated as $l_{loc} \sim v_{k_j}\t(\e_j)$ using Eqs. \ref{GT}, \ref{Te}, \ref{ke} for $j$th mobility 
	edge which results in $l_{loc} \sim 1/\G$. Thus for the same instance of ANT with $n = 12$ we obtain numerically $l_{loc} \sim 400$ nm, therefore such a NT should extend 
	to more than $\sim 5 \,\m$m in length. 
	
	At least, it should be especially noted that, in accordance with the reasoning in Sec. \ref{TNT}, the highest sensibility of NT structures to impurity perturbations and 
	the richest variety of resulting intermittent conductive and localized spectrum ranges in them is expected in the properly designed TNT's at a proper choice of impurity 
	centers and their concentrations.   
	
	\section{Acknowledgements}
	The authors are thankful to L.S. Brizhik, A.A. Eremko, and S.G. Sharapov for their attention to this work and its valuable discussion. V.L. acknowledges the partial 
	support of his work by the Simons Foundation.

	\appendix
	
	\section{Locator and correlator}
	
	We calculate the integral that contributes to the locator GF in a 1D nanosystem, for an example of $\e_k$ mode in ANT:
	\be
	g(\e) = \frac 1{4\pi}\int_{-2\pi}^{2\pi} \frac {dk}{\e - 1 + 2\cos\tfrac k2},
	\ee
	valid at $-3 < \e < 1$. By the common change of variable, $t = \tan \tfrac k4$, this integral is rewritten as:
	\be
	g(\e) = \frac 1\pi\int_{-\infty}^\infty  \frac{dt}{1 + \e - (3 - \e)t^2},
	\ee
	giving the explicit result:
	\be
	g(\e) = i\frac {\theta(1 + \e)\theta(3 - \e)}{\sqrt{(3 - \e)(1 + \e)}}
	\ee
	with the standard $\theta$-functions delimiting the $\e_k$ energy range.
	
	Another contribution to $G_0(\e)$ from $-\e_k$ mode, valid at $-1 < \e < 3$, is
	\be
	g(-\e) = i\frac {\theta(1 - \e)\theta(3 + \e)}{\sqrt{(3 + \e)(1 - \e)}}
	\ee
	which then enters the full expression for locator GF:
	\be
	G_0(\e) = \frac {g(\e) - g(-\e)}{4n}.
	\ee
	
	The next step is to calculate, for the same ANT, the correlator between a pair of impurities at distance $r$: 
	\be
	G_r(\e) = \frac{g_r(\e) - g_r(-\e)}{4n},
	\lb{corr}
	\ee
	through the integral:
	\be
	g_r(\e) = \frac 1{4\pi}\int_{-2\pi}^{2\pi} f_r(k,\e) dk,
	\lb{gre}
	\ee
	\begin{figure}[h]
		\centering
		\includegraphics[width=9cm]{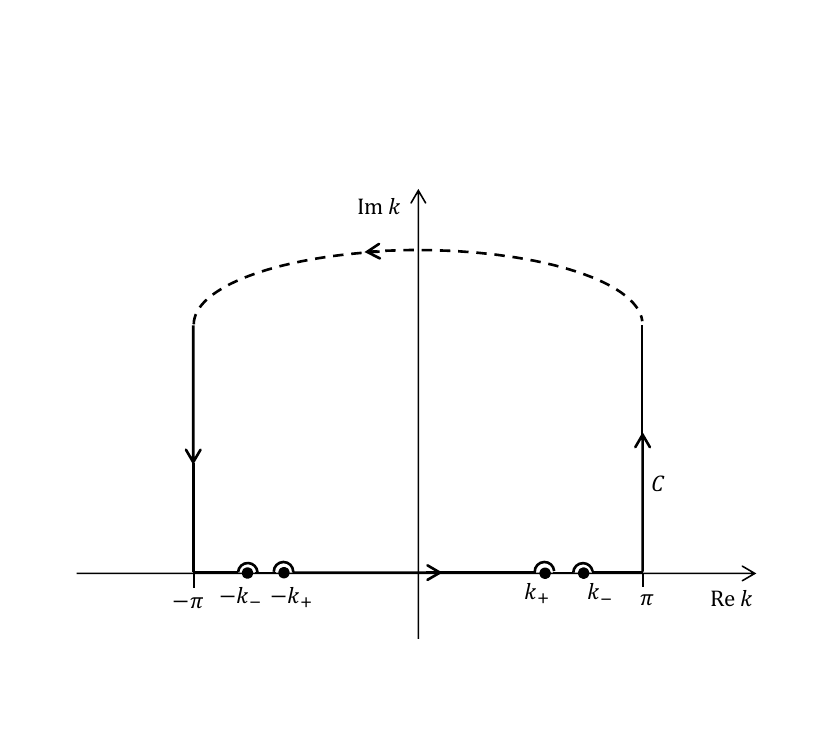}
		\caption{Integration contour for Eq. \ref{cont1}.}
		\label{cont}
	\end{figure}
	with its integrand:
	\[ f_r(k,\e) = \frac{{\rm e}^{ikr}}{\e - 1 + 2\cos\tfrac k2},\]
	especially considering long distances, $r \gg 1$. It can be done passing to the contour integral:
	\bea
	&&\int_C f_r(k,\e)dk = g_r(\e) + \int_0^\infty\left[f_r(2\pi + iy,\e)\right.\nn\\
	&& \qquad\qquad\qquad  \left. - f_r(-2\pi + iy,\e)\right]dy = 0.
	\lb{cont1}
	\eea
	where the contour $C$ in the complex momentum plane (Fig. \ref{cont}) includes the $f_r(k,\e)$ poles:
	\[\pm k_\e = \pm 2\arccos \frac{1 - \e}2.\]
	
	For integration along the imaginary axis we use the relations $\cos(\pm\pi + i y/2) = -  \cosh (y/2)$ and ${\rm e}^{i(\pm2\pi + iy)r} = (-1)^{2r}
	{\rm e}^{-yr}$ (noting that the longitudinal distance $r$ between impurities takes here only integer or half-integer values) and the explicit formula:
	\bea
	&& I_{2r}(b) = \int_0^\infty \frac{{\rm e}^{-2r y}\, dy}{\cosh y + b} = \frac{1}{2r + 1}\left[\left(\frac{b}{\sqrt{b^2 - 1}}   + 1\right)\right.\nn\\
	&&\times {_2F_1}\left(1, 2r + 1; 2r + 2;\frac 1{\sqrt{ b^2 - 1} - b}\right) + \left(\frac{b}{\sqrt{b^2 - 1}}  - 1\right)\nn\\
	&&\times \left.{_2F_1}\left(1,2r + 1;2r + 2;- \frac 1{\sqrt{b^2 - 1} + b}\right)\right]
	\lb{HG}
	\eea
	where ${_2F_1}(n,m;p;q)$ is the hypergeometric function \cite{Andrews}.
	\begin{figure}[h]
		\includegraphics[width=8cm]{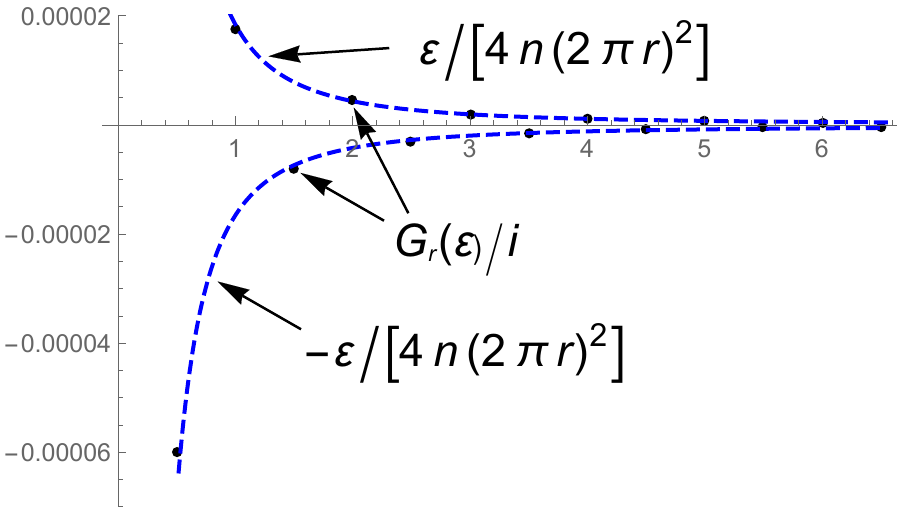}
		\caption{Exact discrete values of the correlator $G_r(\e)$ by Eq. \ref{Gr} (solid points) and their approximation by Eq. \ref{Gapp} (dashed curves) for 
			the choice of $\e = \e_{res}$ and $n = 12$.}
		\label{figs} 
	\end{figure}
	Then the sought correlator follows from Eqs. \ref{corr}, \ref{gre}, \ref{HG} analytically as:
	\bea
	&& G_r(\e) = \frac{(-1)^{2r}}{16\pi n}\left[I_{2r - 1}\left(b_\e\right) - I_{2r + 1}\left(b_\e\right)\right.\nn\\
	&&\qquad\qquad\quad \left. - \,I_{2r - 1}\left(b_{-\e}\right) +  I_{2r + 1}\left(b_{-\e}\right)\right]
	\lb{Gr}
	\eea
	with the energy dependent parameter
	\[b_\e = \frac {1 - \e}2.\]
	Notably, the full form by Eq. \ref{Gr} admits a very simple approximation:
	\be
	G_r(\e) \approx i\frac{(-1)^{2r}\e}{4 n (2\pi r)^2},
	\lb{Gapp}
	\ee
	however quite precise at all non-zero inter-impurity distances (see Fig. \ref{figs}) and suitable for detailed evaluations as in analysis of particular GE 
	terms (Sec. \ref{Beyond}B).

\end{document}